\documentclass[pre,aps,preprint]{revtex4-1}
\pdfoutput=1
\usepackage{amsmath,natbib,amsfonts}
\usepackage{amssymb,color}
\usepackage{graphicx}
\newcommand{\beqa}{\begin{eqnarray}}
\newcommand{\eeqa}{\end{eqnarray}}

\begin{document}

\title{Stochastic Work Extraction in a colloidal heat engine in presence of colored noise}
\author{Arnab Saha}
\email{sahaarn@gmail.com}
\affiliation{Department of Physics, Savitribai Phule Pune University, Ganeshkhind, Pune  411007, India.}
\author{Rahul Marathe}
\email{maratherahul@physics.iitd.ac.in}
\affiliation{Department of Physics, Indian Institute of Technology, Delhi, Hauz Khas 110016, New Delhi, India.}

\date{\today}
\begin{abstract}
From synthetic active devices such as self-propelling Janus colloids to micro-organisms like bacteria, micro-algae, living cells in tissues, active fluctuations are ubiquitous. Thermodynamics of small     systems involving thermal as well as active fluctuations are of immense importance. They can be    employed to extract thermodynamic work. Here we propose a simple model system that can produce thermodynamic work exploiting active fluctuations. We consider a Brownian particle, trapped by an externally controlled harmonic confinement that contracts and expands time-periodically by modulating its spring constant e.g an optical tweezer. The system produces work by being alternately connected  to two baths one passive and other active. The active bath provides exponentially time-correlated noise to the particle, that breaks the fluctuation dissipation relation. The average efficiency of the system is calculated exactly in quasistatic limit. Nonquasistatic regime is explored by numerics. Comparing with its passive counterpart, we also show that the active micro heat engine can be more efficient depending on the chosen parameter space. We also believe that our model can be realised experimentally with the help of bacterial baths.     
\end{abstract}
\maketitle

\section{Introduction}

It is demonstrated in the seminal contribution of Wu and Libchaber that diffusivity of a micron scale colloidal bead in a freely suspended soap film containing a population of {\it {Escherichia coli}} in a quasi two dimensional set up, is distinct from the diffusivity of a Brownian particle, at least for short time i.e. below 10 seconds \cite{Libchab2000}. The bead is super diffusive in short times indicating persistent fluctuations in its dynamics. The origin of such anomalous behaviour is attributed to the incessant collisions of the bead with the bacteria forming emergent patterns in the soap film. The life time of such patterns depends how long the bead can be super diffusive. This study motivated several others to investigate transport properties and rheology of active media which is fundamentally different from their passive counterpart primarily because of active, out-of-equilibrium fluctuations. Some of the aspects studied are enhanced diffusion/reduction of viscosity in presence of motile bacteria, directed cellular transport, characterisation of effective viscosity of microswimmer suspensions, rheology of soft materials etc. \cite{kim2004, caspi2002, rafai2010, chen2010, sokolov2009}.    

Recently, the investigation on small scale thermodynamics of the colloidal bead suspended in an active medium such as bacterial film or solutions has drawn much attention. One of the major motivations comes from thermodynamic work extraction. Even in macroscopic thermodynamics, arguably the most important application of fundamental principles is thermal machines, such as heat engines \cite{Callen}. Recently, with the advent of novel technologies and complementary support from theory of small scale thermodynamics (namely, stochastic thermodynamics \cite{Sekimoto97, Seifert12}), the miniaturisation of macro heat engines at the scales of a single colloidal particle have become possible. For example, microscopic heat engines with a trapped, time-periodically driven colloidal particle, immersed in passive aqueous solution and subjected to Stirling as well as Carnot protocols have now been realised experimentally \cite{bechinger10, Martinez16, Edgar16}. The key ingredients behind the experimental as well as theoretical techniques to construct such micron-scale thermal machines is their capability to take into account the out-of-equilibrium fluctuations, which are predominantly present in the dynamics of small world. Now one can proceed further together with these frontier techniques to explore the possibilities of extracting thermodynamic work exploiting the dynamics of a colloidal bead under time-dependent confinement and {\it{active fluctuations}} present in the suspensions of living organisms, such as live bacteria, micro-algae etc. or synthetically active Janus colloids. Recently it has been explored in experiments by driving a harmonically trapped colloidal particle, suspended in a bacterial bath, with Stirling protocol \cite{Sood16}. Where it has been shown that the efficiency of such a micro-machine can considerably be high, thanks to the non-Gaussian nature of active fluctuations involved. However, the issue of the enhancement in the efficiency in presence of active fluctuations is a matter of current debate \cite{Zakine17}.

Earlier it has been demonstrated experimentally that active, self-propelling entities such as live bacteria or micro-algae can affect the fluid viscosity and thereby the dissipation within the fluid,  depending on the concentration and swim speed of the suspended microorganisms \cite{kim2004, caspi2002, rafai2010, chen2010, sokolov2009}. Motivated by this observation, one can constitute a theoretical model of an active micro heat engine in presence of {\it{non-equilibrium heat bath}} with active dissipation \cite{Saha18}.    

Here, we consider a harmonically confined colloidal particle as a working substance. The trap strength (or the spring constant of the trap) follows a time dependent protocol popularly known as the Stirling protocol \cite{bechinger10}, mimicking different strokes in an engine. The full protocol takes time say $\tau$. In the first half of the protocol, the particle is in contact with a thermal bath at temperature $T$, modelled by a Gaussian white noise which follows the fluctuation dissipation relation (FDR). In this half of the cycle, we assume that the activity due to suspended organisms is negligible and has no effect on the bath characteristics. In the second half of the protocol, the activity of micro-organisms is increased (e.g by adding food, or flashing light \cite{Sood16}). The effect of the enhanced activity alters the properties of the heat bath which in turn alters both the characteristics of the noise and the dissipation constant. Theoretically this can be achieved by an active Ornstein Uhlenbeck process (AOUP) \cite{MaggiPRL14, Fodor16, MandalPRL17, Cates18, Vulpiani19, Rajarshi19}. The AOUP breaks the FDR and models the persistent dynamics of a colloidal particle suspended in an active, non-equilibrium heat bath. The dynamics of the colloidal particle is taken to be persistent up to a certain correlation time $\tau_a$. In AOUP noise is exponentially correlated within the time scale $\tau_a$ (check section \ref{an_results} for details). Breaking of FDR with finite $\tau_a$ signifies finite activity that drives the bath out of equilibrium whereas restoring FDR with $\tau_a \rightarrow 0$ limit gives the thermal bath equilibrated at temperature $T$. For finite $\tau_a$ thermodynamic temperature of the bath is undefined, but with the help of stochastic thermodynamics, one can still define thermodynamic work, heat and therefore the efficiency of such an engine. For the model considered here we analytically find average work and efficiency in the quasistatic limit (large protocol times $\tau$). At small protocol times (small $\tau$), the system is nonquasistatic. We explore this regime with simulation. Finally, we compare the average thermodynamic efficiency of our model with passive, microscopic, colloidal Stirling engine, which was experimentally implemented in \cite{bechinger10}. In particular, we show that depending on the chosen parameter space, the active heat engine can be more efficient than its passive counterpart.
     
We also note that recently particles following AOUP promisingly manifest important collective phenomena such as, motility induced phase separation in two dimensions \cite{Cates15} which is common in real active systems such as self-propelling colloids \cite{Fodor16}. It should also be mentioned here that there exist other important particle or agent based theoretical models for active systems (e.g. active Brownian motion \cite{zottl2016}, run and tumble dynamics \cite{tailleur2008, elgeti2015run,soto2014run}, velocity depot/velocity dependent friction models etc. \cite{romanczuk2012depo}) demonstrating various individual as well as collective properties of real active systems. Equivalence among all such models is an issue of current debate \cite{cates2013equi}. However, extension of stochastic thermodynamics to such model systems may open up myriad possibilities. 

The paper is organized as follows. In section \ref{model} we describe the model, and thermodynamic quantities of interest.  In section \ref{an_results} we discuss the
most general case and calculate analytically all the thermodynamic quantities in the quasistatic limit. We also simulate the system in the non-quasistatic regime as well
as quasistatic regime. In section \ref{part_results}, we consider a particular case of our model where we find analytical expressions in the quasistatic case. 
We also simulate our model and compare it with the analytical results. Finally we conclude with a discussion in section \ref{conclude}.


\section{Model} \label{model}
\begin{figure}[!t]
\centering
\includegraphics[width=11cm, height=7cm, angle=0]{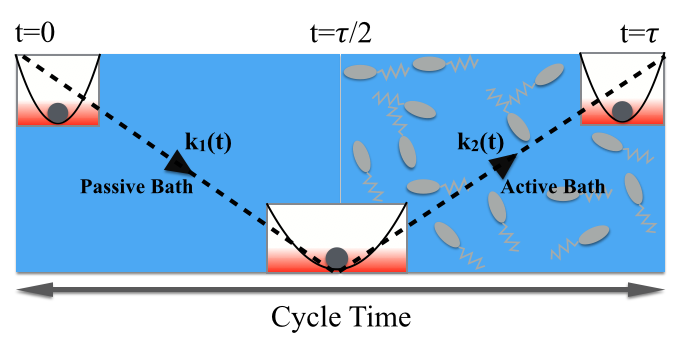}
\caption{Cartoon of the time variation of the trap strength and the activity of the heat bath. First half corresponds to the passive branch (isothermal expansion step) and second half is the active branch (compression step) as indicated in the figure. At the half of the cycle $t=\tau/2$ when $k(t)$ is at it's minimum, activity is switched {\it on} for the full second half of the cycle and then it is switched {\it off} at $t=\tau$, such that the system returns to it's original state.}
\label{prot_fig}
\end{figure}

In this section we describe the model in detail. We consider a one dimensional system that consists of a colloidal particle in contact with a reservoir and is confined within a parabolic potential. The strength of the trap together with the bath properties vary periodically in time. This drives the particle out of the equilibrium. The trap-strength $k(t)$ varies within a time-period $\tau$.
\beqa
k(t)&=&k_{max}+(k_{min}-k_{max}) \frac{2t}{\tau} ~~~~~~~~=k_1(t),~~~~0\leq t\leq\tau/2 \nonumber \\ 
&=&k_{min}+(k_{min}-k_{max})(1-\frac{2t}{\tau}\ )=k_2(t),~~~~\tau/2 < t\leq\tau.
\label{protocol1}
\eeqa
Here $k(t)$ varies linearly between $k_{max}=k_0$ and $k_{min}$ within a cycle, as shown in figure (\ref{prot_fig}). In general we may have $k_{min}=k_0/n$ (where $n\ge2$).

First we use $n=2$ to derive average efficiency of the system. However, later we discuss the generalisation of the protocol and it's consequences on the efficiency of our micro-machine. The whole assembly i.e. colloidal particle plus trap is suspended in an aqueous medium consisting of active particles. The activity of the bath is also turned on and off repeatedly in a time periodic manner. For instance in the first half of a cycle (i.e. $0\leq t\leq\tau/2$) the bath is passive i.e. the particles of the bath are undergoing only thermal motion at temperature $T$. In the second half (i.e. $\tau/2 < t\leq\tau$), an external agent switches on the activity (e.g. by adding food, or shining light \cite{Sood16}) of the active particles hence altering the properties of the heat bath. In general, both the frictional drag on the colloidal particle due to its surrounding fluid and the time-correlation between the random forces on the colloidal particle due to its collisions with bath particles are modified when the bath becomes active from passive and vice versa. This turning on and off the bath activity is repeated over several cycles such that the system reaches a non-equilibrium steady state. The cycle time $\tau$ determines whether the drive is quasistatic or not. If $\tau$ is very large compared to any other time scale involved, the drive from the trap of the harmonic confinement is very slow and thereby the dynamics is quasistatic, otherwise it is nonquasistatic. Decrease in the trap strength from $t=0$ to $t=\tau/2$ corresponds to isothermal expansion in contact with the passive bath equilibrated at temperature $T$. Switching on the activity at $t=\tau/2$, and then increase in trap strength from $t=\tau/2$ to $t=\tau$ corresponds to compression in presence of an active bath. Due to activity the bath is out of equilibrium and therefore temperature cannot be defined. Finally at $t=\tau$ activity is switched off and the cycle in completed.  These are the four strokes of our micro heat engine. We believe such a system may be realised experimentally using photo active bacteria or Janus particles with a periodic on-off switching of a light source.

For such a system the dynamics of the position of the colloidal particle $x(t)$ can be modeled by an overdamped Langevin equation as,  
\beqa
\gamma_1 \dot{x}&=& -k_1(t)x+\sqrt{D_1}~\xi_1(t), ~~~~~~~0\leq t\leq\tau/2.
\label{Lang1}
\eeqa
Here, $\gamma_1$ is the friction coefficient, thermal noise strength $D_1=2\gamma_1 k_BT$ where $k_B$ is the Boltzmann constant, $T$ is the temperature of the passive, thermal reservoir. The random forces due to the collisions between the colloidal particle and the bath particles are modeled by $\xi_1(t)$, which is a Gaussian white noise following $\langle\xi_1(t)\rangle=0$ and $\langle\xi_1(t)\xi_1(t^{\prime})\rangle=\delta(t-t^{\prime})$. Clearly, during the first half of the cycle, fluctuation dissipation relation (FDR) between the friction and the noise strength is maintained. Therefore for a time-independent trap strength, the colloidal particle will relax to an equilibrium state at temperature $T$.

In the second half of the cycle, the reservoir becomes active due to the presence of active entities (e.g. bacteria). To model active particles we use active Ornstein-Uhlenbeck process (AOUP) \cite{MaggiPRL14, Fodor16, MandalPRL17, Cates18, Marconi16}. For this process corresponding Langevin equation becomes,
\beqa
\gamma_2 \dot{x}&=&-k_2(t)x+ \sqrt{D_2}~\xi_2(t), ~~~~~~~\tau/2 < t\leq\tau,
\label{Lang2}
\eeqa
and
\beqa
\tau_a\frac{d\xi_2(t)}{dt}&=&-\xi_2(t)+\sqrt{\tau_a}~\xi_1(t), ~~~~~~~\tau/2 < t\leq\tau.
\label{Lang3}
\eeqa
First, we note here that as activity is switched on in the second half of the cycle, friction coefficient of the particle is altered to $\gamma_2$. Second, due to activity, the noise, which was $\xi_1(t)$ becomes $\xi_2(t)$ which follows a dynamics given by Eq. (\ref{Lang3}). Solving Eq. (\ref{Lang3}) it is easy to show that the {\it active noise}, $\xi_2(t)$ is exponentially correlated,
\beqa
\langle\xi_2(t)\rangle=0;  ~~~~~~
\langle\xi_2(t)\xi_2(t^{\prime})\rangle=\frac{1}{2}\ e^{-\frac{|t-t^{\prime}|}{\tau_{a}}},
\label{xi2corr}
\eeqa 
where $\tau_{a}$ is the finite correlation time scale of the active noise. In general, the strength of the active noise $\sqrt{D_2}$ and the friction coefficient $\gamma_2$ are independent and are not related by FDR. Furthermore, one can note here that the active noise-noise correlation (Eq. (\ref{xi2corr})) has a memory unlike the dissipative term $\gamma_2 \dot{x}$ in Eq. (\ref{Lang2}) which is memoryless. Therefore, in the second half of the cycle FDR is broken and the bath is driven out of equilibrium.

We are interested in quantifying stochastic thermodynamic quantities, such as work done, heat exchanged between the system and the bath along a cycle, efficiency etc. and their averages over trajectories in the steady state. From Eqs. (\ref {Lang1}) and (\ref{Lang2}), using Stratonovich convention, we obtain $dq=(\sqrt{D_i}\xi_i-\gamma_i\dot{x})\dot{x}dt=\frac{d}{dt}\left(\frac{1}{2}k_ix^2\right)dt-\left(\frac{1}{2}\dot{k_i}x^2\right)dt=du-dw$, ($i\in (1,2)$) where $du, dw, dq$ are identified as infinitesimal change in internal energy, infinitesimal work done and infinitesimal heat exchanged between the system and the bath along a trajectory, within infinitesimally small time interval $dt$. This is known as the {\it {first law}} in stochastic thermodynamics \cite{Sekimoto97}. According to our sign convention the work done on the system and heat absorbed by the system are positive. One can calculate $dw$ and $du$ along a trajectory of the particle, following the definitions above. Then the first law of stochastic thermodynamics can be applied to calculate $dq$ instead of the definition mentioned before. Next, one can integrate the thermodynamic quantities, in particular the thermodynamic work along a trajectory. To determine average values, we average over all possible noise realisations. We determine average efficiency $\eta$ of the system we follow,
\beqa
\eta=\frac{-\langle W\rangle }{\langle Q_{in}\rangle}
\eeqa          
where $\langle W\rangle $ is the average work done by the system through out a cycle and $\langle Q_{in}\rangle $ is the average heat input to the system i.e. the heat exchanged between the system and the bath along the isothermal expansion by the protocol $k(t)$ within $0\leq t\leq \tau/2$.  We calculate all the thermodynamic quantities including efficiency of the system in nonequilibrium steady state.  

In quasistatic limit, results are obtained analytically as well as from simulations. We then compare the analytical and numerical results. We obtain nonquasistatic results from simulations with small $\tau$. Here the quasistatic limit corresponds to the cycle time $\tau $ much greater than any other time scale involved e.g. $(\gamma_1^{-1}, \gamma_2^{-1})\sim 1 $ or the active correlation time $\tau_a\sim 10$ (in units of $\gamma_2^{-1}$). In simulations we integrate the equations of motion Eqs. (\ref{Lang1}) and (\ref{Lang2}) by a velocity Verlet algorithm using Stratonovich discretization with time step $dt\sim 10^{-3}$ and obtain average work and heat exchanged. These averages are taken over $10^5$ cycles of $k(t)$, after driving the system in the steady state. In simulations we choose the mass of the particle to be unity and the units of time, length and energy are $\gamma_2^{-1}$, $\sqrt{\frac{k_BT}{k_0}}$, $k_BT$ respectively.

\section{General Case} \label{an_results}

Here we discuss the generic case where, due to activity, both fluctuation and dissipation alters breaking FDR, as mentioned before. We calculate average thermodynamic quantities like work, heat and efficiency analytically in the quasistatic limit. We begin by writing down the dynamics of position fluctuation $\sigma_x=\langle x^2\rangle$ of the particle derived from Eqs. (\ref{Lang1}) and (\ref{Lang2}). We multiply these equations by $x(t)$ and take average to obtain,
\beqa
\gamma_i\frac{d\sigma_x(t)}{dt}=-2k_i\sigma_x(t)+2\sqrt{D_i}~\langle \xi_i(t) x(t)\rangle ,
\label{dsigma}
\eeqa    
where $i\in(1,2)$. This equation is true for all cycle times. One can solve Eq. (\ref{dsigma}) numerically to get $\sigma_x(t)$. We may also use the formal solution of equations of motion Eqs. (\ref{Lang1}) and (\ref{Lang2}) for this purpose,
\beqa
x(t)=\exp\left(-\frac{1}{\gamma_i}\int^tk_i(t^{\prime})dt^{\prime}\right) \left(x_{init.}+\frac{\sqrt{D_i}}{\gamma_i}\int^tdt^{\prime}~\xi_i(t^{\prime})~\exp\left(\frac{1}{\gamma_i}\int^{t^{\prime}}k_i(t^{\prime\prime})~dt^{\prime\prime}\right)\right).
\label{xt}
\eeqa 
Substituting particular form for $k_1(t)$ with $k_{max}=k_0$ and $k_{min}=k_0/2$ in Eq. (\ref{protocol1}). For first half of the cycle, Eq. (\ref{xt}) gives,
\beqa
x(t)&=&  \exp\left(-\frac{k_0}{\gamma_1}\ (t-\frac{t^2}{2\tau}\ )\right) 
\left(x_0 + \frac{\sqrt{D_1}}{\gamma_1}\ \int_0^t \exp\left(\frac{k_0}{\gamma_1}\ (t'-\frac{t'^2}{2\tau}\ )\right)~ \xi_1(t') ~dt'\right). 
\eeqa
Multiplying above equation by $\xi_1(t)$ and taking average over noise and initial conditions we get,
\beqa
\langle x(t)\xi_1(t) \rangle = \frac{\sqrt{D_1}}{2\gamma_1}.
\label{passeq1}
\eeqa
Similarly for the second half of the cycle with $k_2(t)$ in Eq. (\ref{protocol1}), we multiply, 
\beqa
x(t)&=&  \exp\left(-\frac{k_0}{2\gamma_2\tau}\ (t^2-(\tau/2)^2) \right) 
\left(x_{\tau/2} + \frac{\sqrt{D_2}}{\gamma_2}\ \int_{\frac{\tau}{2}}^t \exp\left(\frac{k_0}{2\gamma_2\tau}\ (t'^2-(\tau/2)^2) \right)~ \xi_2(t') ~dt'\right),\nonumber\\
\label{xtapp}
\eeqa
with $\xi_2(t)$ and take average over the exponentially correlated noise. Using $\langle \xi_2(t)\xi_2(t')\rangle = \frac{1}{2}\ e^{-|t-t'|/\tau_a}$ we get,
\beqa
\langle x(t)\xi_2(t) \rangle = \frac{\sqrt{D_2}}{2\gamma_2}\ \exp\left(-\frac{k_0}{2\gamma_2\tau}\ t^2 -\frac{t}{\tau_a}\ \right) 
\int_{\tau/2}^t \exp\left(\frac{k_0}{2\gamma_2\tau}\ t'^2+\frac{t'}{\tau_a}\ \right)~dt'. 
\eeqa
By completing the squares in the integral above and defining $\alpha=\frac{\gamma_2 \tau}{k_0\tau_a}\ $,
$\lambda = \frac{k_0}{2\gamma_2\tau}$, we arrive at,
\beqa
\langle x(t)\xi_2(t) \rangle = \frac{\sqrt{D_2}}{2\gamma_2}\ \exp\left(-\left(t+\alpha \right)^2\right)
\int_{\frac{\tau}{2}+\alpha}^{t+\alpha} \exp(\lambda t'^2)~dt'.
\label{xtzt1}
\eeqa
Eq. (\ref{xtzt1}) may be written in terms of error functions, but it is not very illuminating hence we keep it in the form above. We also note that these expressions are 
true even for a non-quasistatic process. 
\begin{figure}[!t]
\centering
\includegraphics[width=0.32\columnwidth]{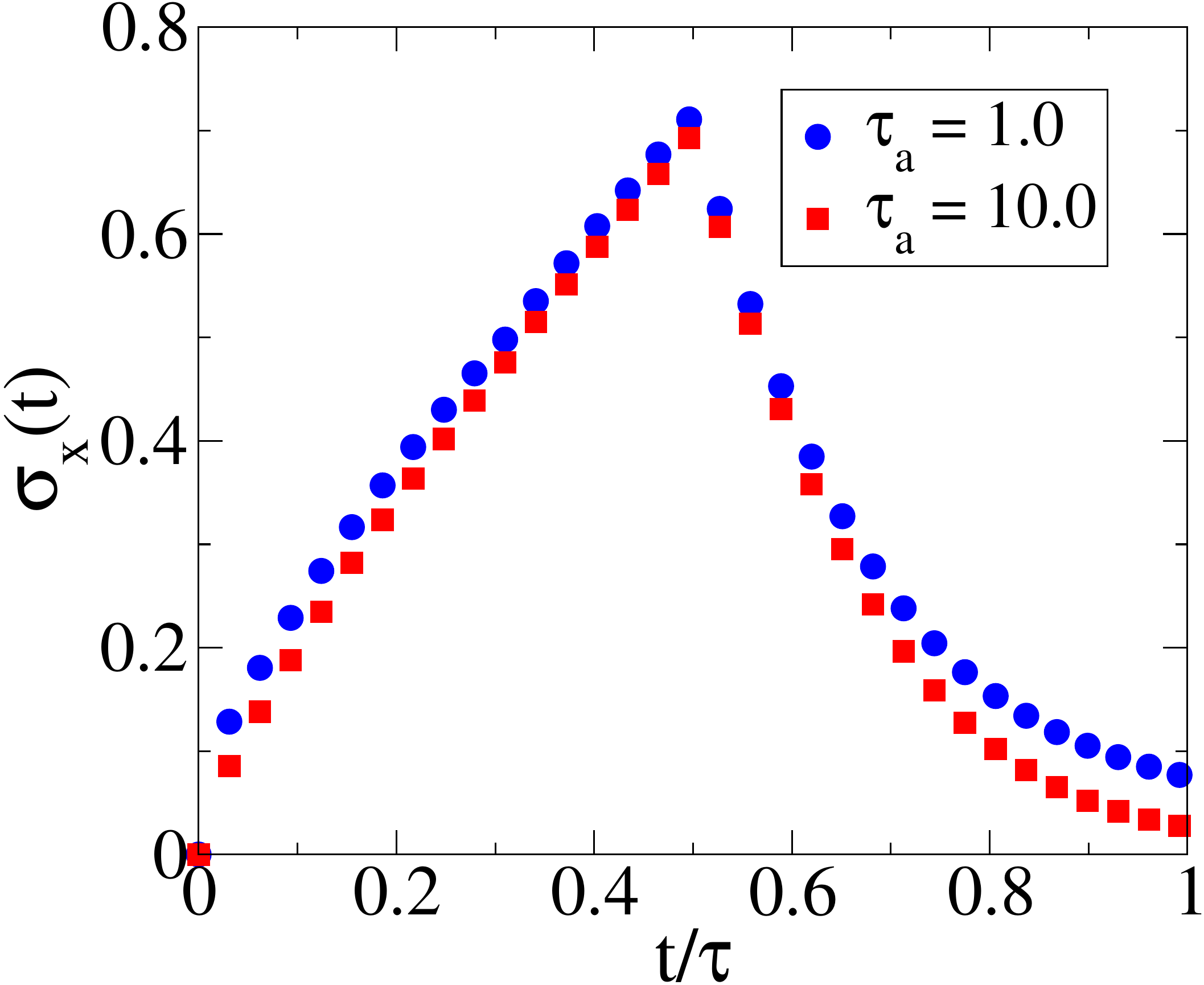}
\includegraphics[width=0.32\columnwidth]{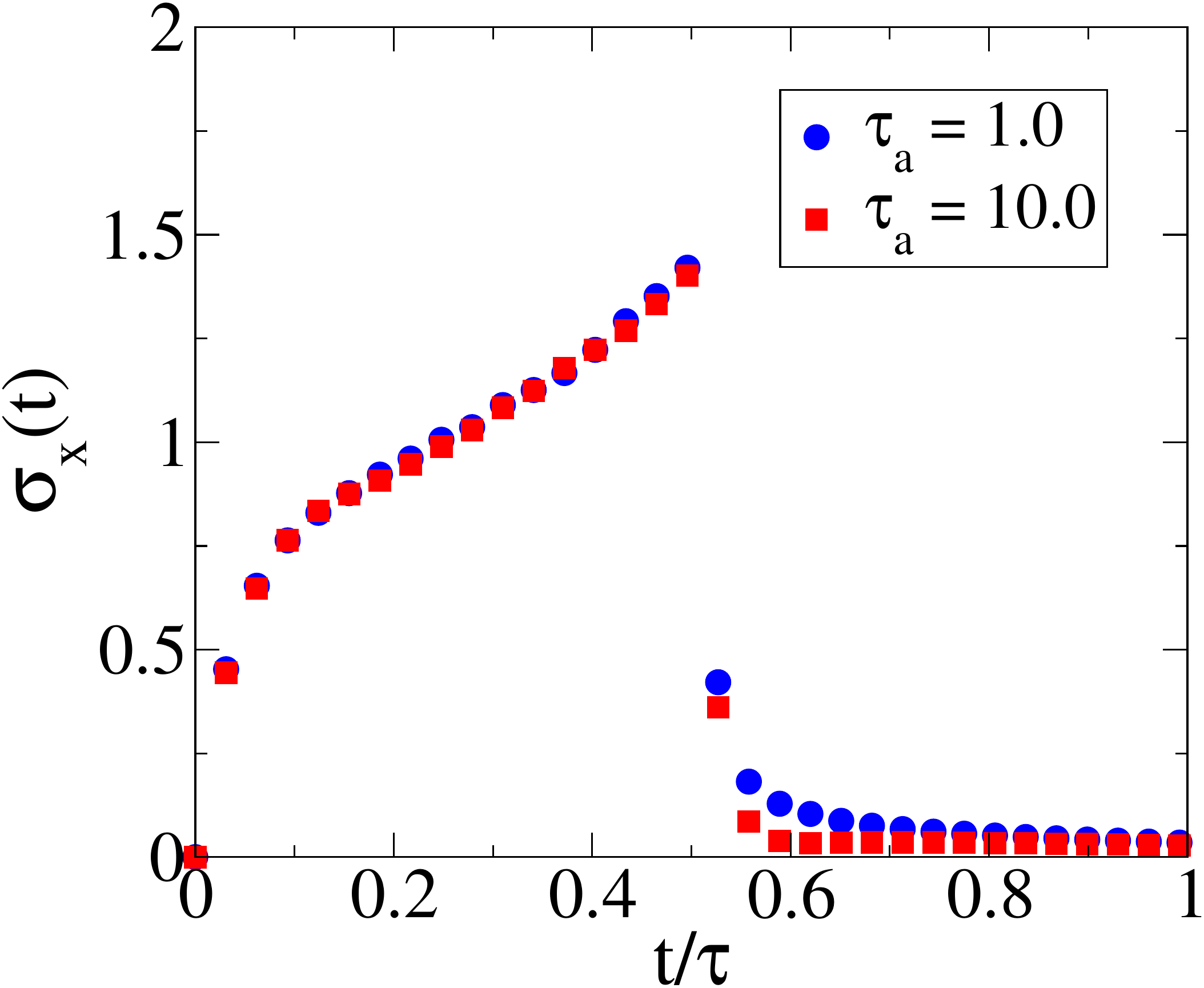}
\includegraphics[width=0.32\columnwidth]{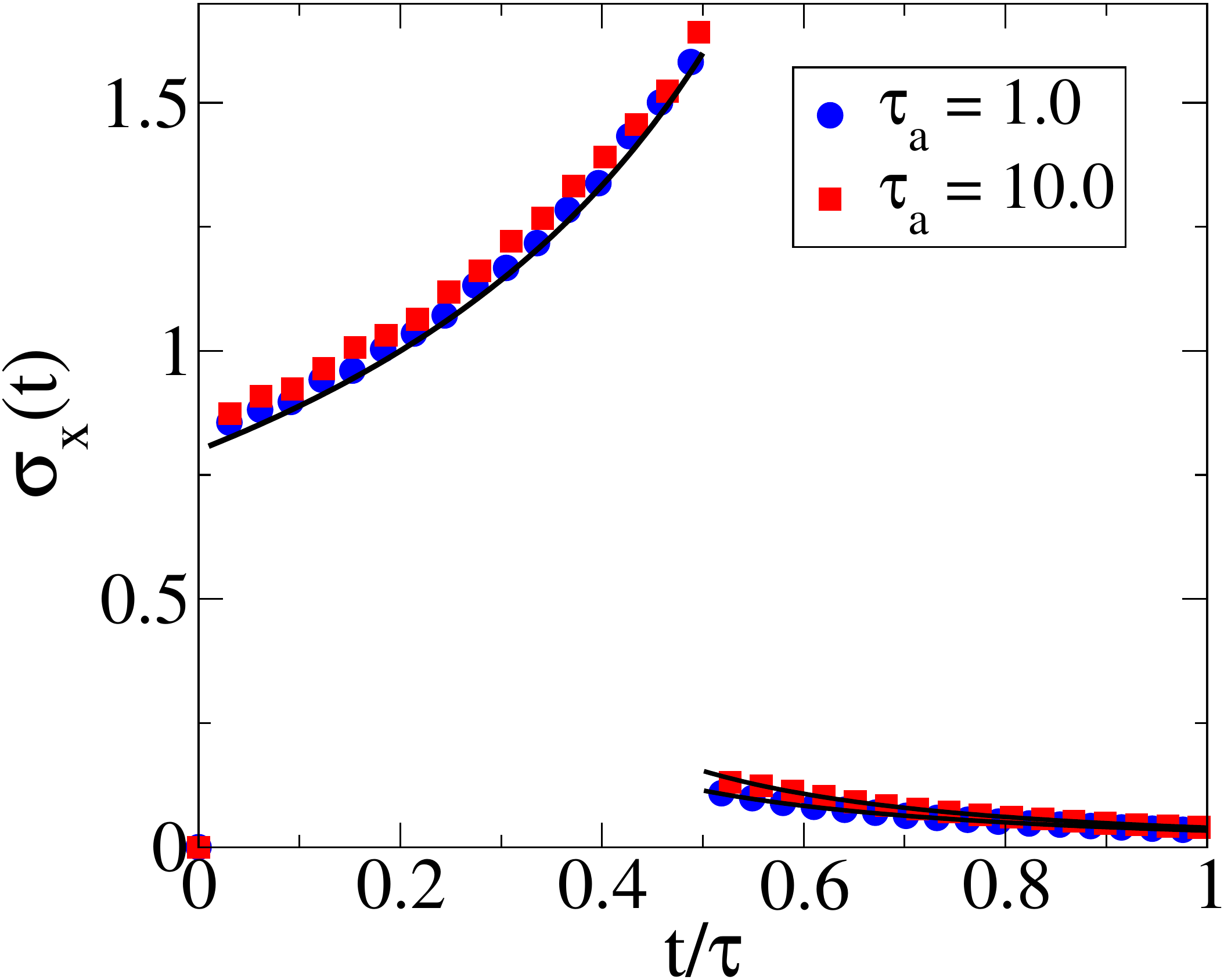}
\caption{Plot of $\sigma_x(t)$ vs $t/\tau$ for different cycle times $\tau=1$ (left panel), $\tau=10$ (middle panel) and $\tau=500$ (right panel) from simulations. Note that for $\tau=500$ (right panel) solid lines correspond to quasistatic analytic expressions in Eqs. (\ref{sigma}). Other parameters are $T=4$, $k_0=5$, $\gamma_1=4$, $\gamma_2=1$, $D_2=5$, $n=2$.}
\label{gensigxt}
\end{figure}
We now turn our attention towards $\sigma_x(t)$ in the long time or quasistatic limit. It will be useful to get analytical expressions for other thermodynamic quantities mentioned earlier. To do this we first calculate correlation between the noise and the position of the trapped particle in two halves of the cycle. We already have the expression for slow expansion of the trap in Eq. (\ref{passeq1}).

\begin{figure}[!t]
\centering
\includegraphics[width=0.32\columnwidth]{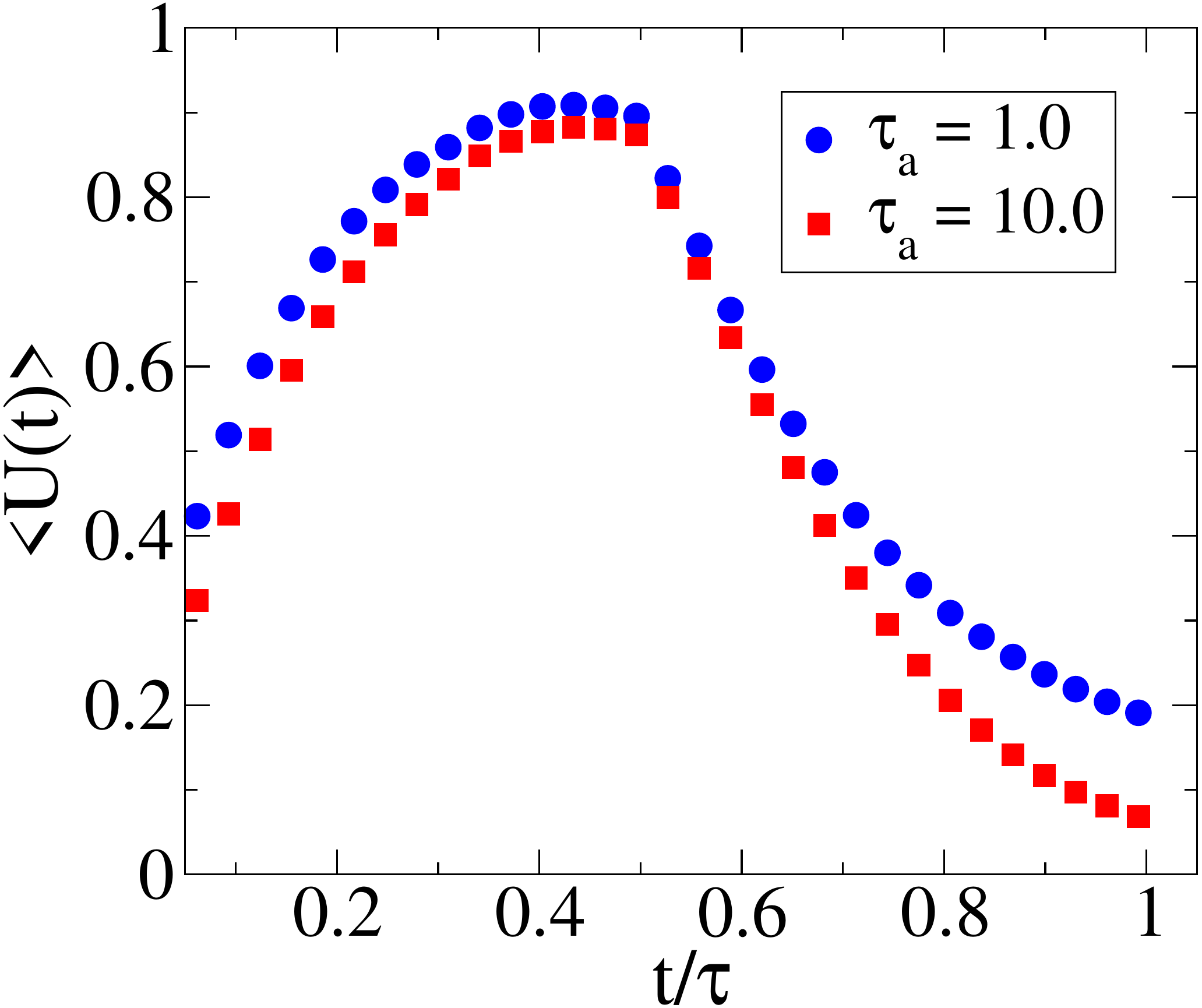}
\includegraphics[width=0.32\columnwidth]{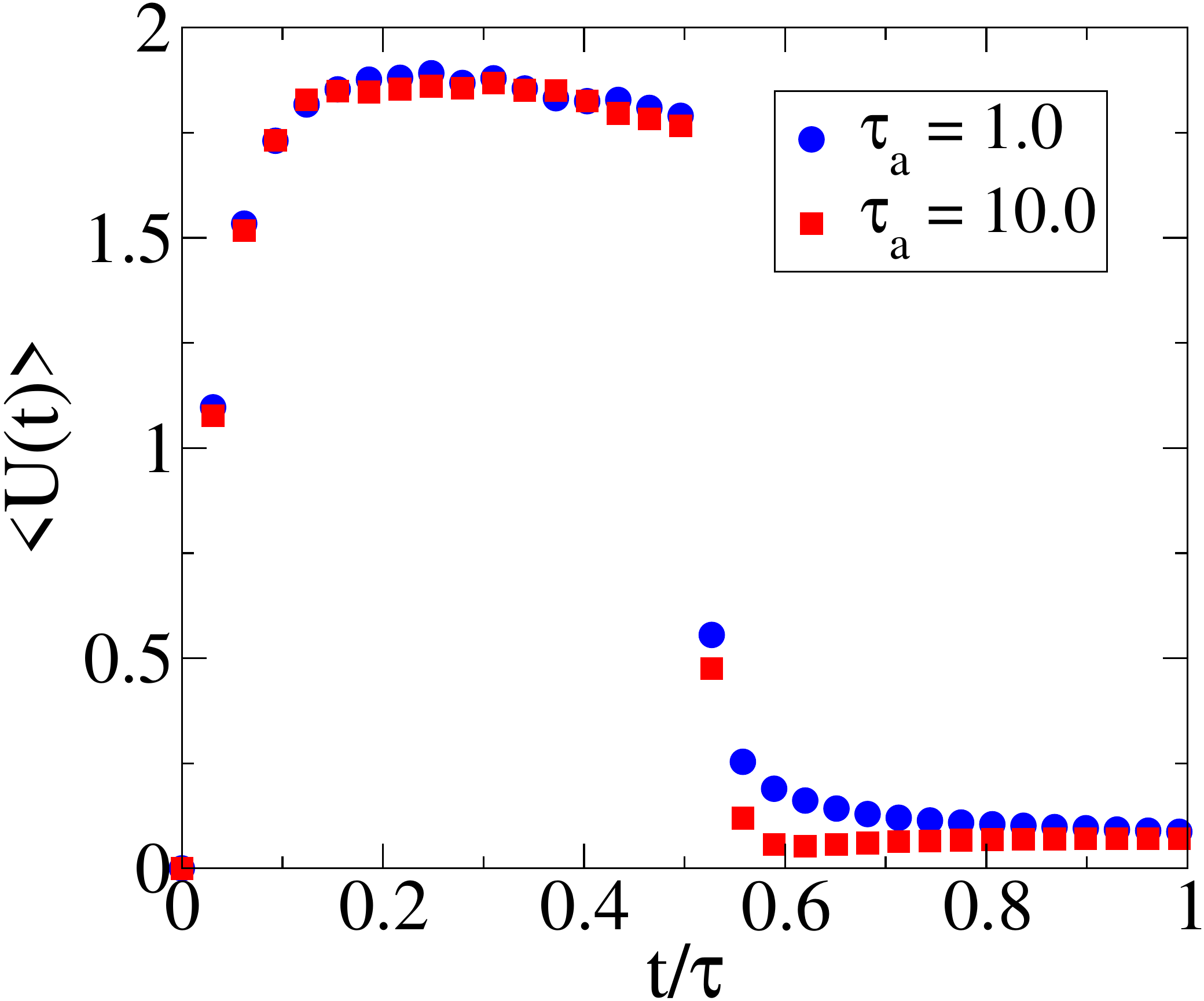}
\includegraphics[width=0.32\columnwidth]{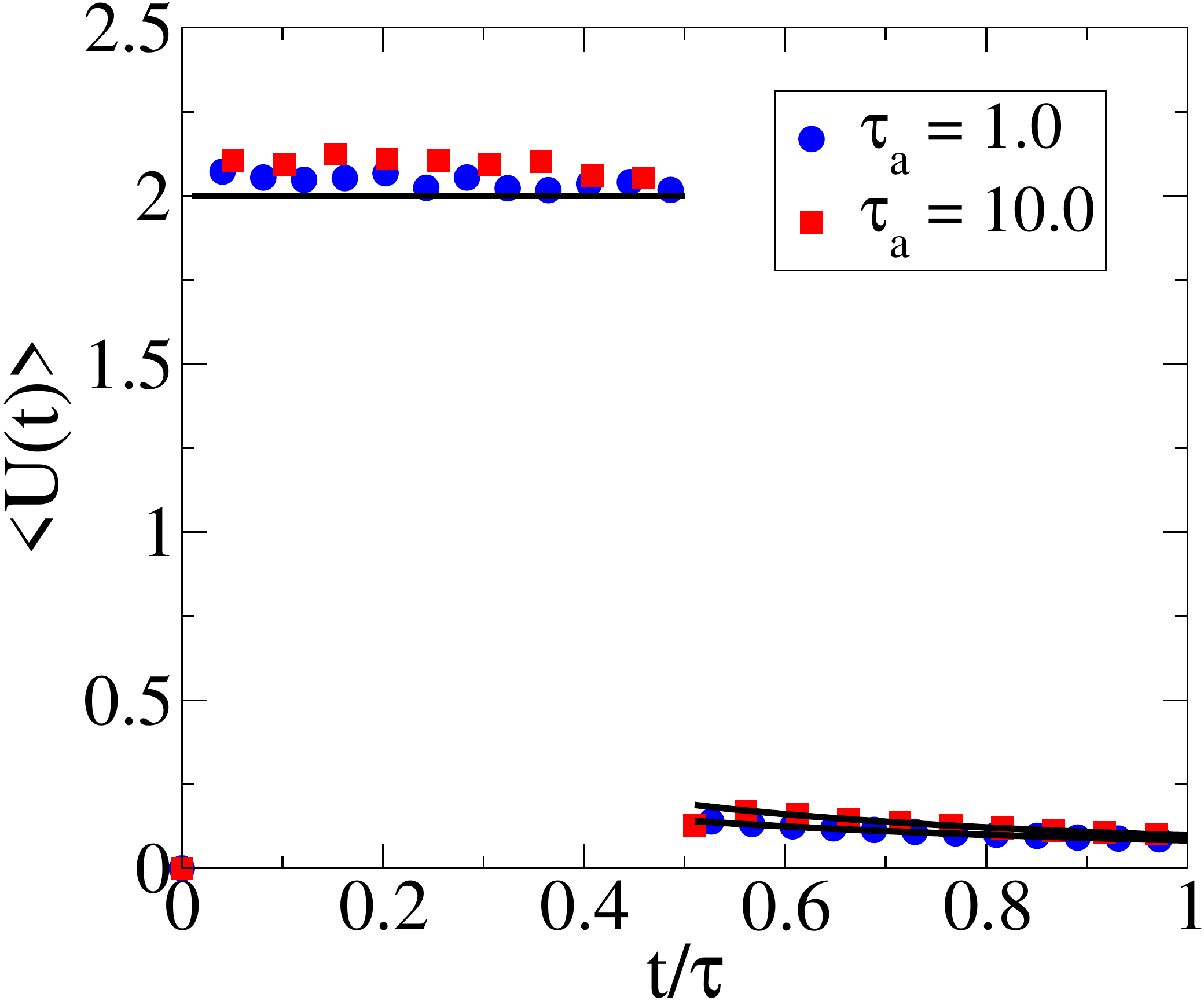}
\caption{Plot of average internal energy $\langle U(t)\rangle $ vs $t/\tau$ for different cycle times $\tau=1$ (left panel), $\tau=10$ middle panel and $\tau=500$ (right panel) from simulations. For $\tau=500$ (right panel) solid lines correspond to quasistatic calculations discussed in the text. Other parameters are $T=4$, $k_0=5$, $\gamma_1=4$, $\gamma_2=1$, $D_2=5$, $n=2$.}
\label{genenergyU}
\end{figure}

For the slow compression step, assuming $k_2(t)$ is a slowly varying function for $\tau >>1$, such that $\int_{\tau/2}^t k_2(t') dt' \sim k_2 (t-\frac{\tau}{2})$ we calculate $\langle x^2(t) \rangle $. After some straight forward algebra and neglecting non-contributing terms in the quasistatic limit we arrive at, 
\beqa
\langle \xi_2(t)x(t)\rangle=\frac{\sqrt{D_2}~\tau_a}{2(\gamma_2 + k_2(t)\tau_a)}\ .
\label{acteq}
\eeqa  
Using above expressions of correlations and from Eq. (\ref{dsigma}) neglecting the term on the left hand side due to quasistaticity, the position fluctuation of the trapped particle becomes,
\beqa
\sigma_x(t) \approx
\begin{cases}
\frac{D_1}{2\gamma_1 k_1(t)}\ , ~~~~~~~~~~~~~~0 < t\leq \tau/2, \\ 
\frac{D_2\tau_a}{2k_2(t)(\gamma_2+k_2(t)\tau_a)}\ , ~~~~\tau/2 < t \leq \tau.
\end{cases}
\label{sigma}
\eeqa
Quasistatic limit is obtained by taking $\tau \rightarrow \infty$, is based on the assumption that the time taken by the position fluctuations of the particle to relax to the values obtained from Eqs. (\ref{sigma}) at a given time point $t$ is negligibly small. Quasistatic limit in similar systems was also explored to calculate relevant thermodynamic quantities analytically in \cite{Saha18, Arun14, ArunPhysicaA}. We note that in the second half of the cycle it is not possible to define an effective temperature $T_{eff}$ since the heat bath is out of equilibrium. In Fig. (\ref{gensigxt}) we plot $\sigma_x$ for different cycle times. We see the quasistatic results match extremely well with the analytical expression in Eq. (\ref{sigma}) for $\tau=500$ in the right panel. While calculating the thermodynamic quantities such as work and heat in quasistatic limit, we will use respective cases in Eq. (\ref{sigma}). For instance, the average energy of the trapped particle is given by $\langle U\rangle =\frac{1}{2}k\sigma_x $. The average energy at large $\tau$ thus becomes $\langle U\rangle =\frac{D_1}{4\gamma_1}$ for the expansion step and $\langle U\rangle =\frac{D_2\tau_a}{4(\gamma_2+k_2(t)\tau_a)}$ for the compression step. In Fig. (\ref{genenergyU}) we plot the average internal energy for different cycles times. We see that the quasistatic case right panel $\tau=500$, matches extremely well with these analytical expression.

Next, using the quasistatic expression of $\sigma_x$, we calculate average thermodynamic work in  this limit. The average work during expansion is given by, 
\beqa
\langle W_1\rangle=\frac{1}{2}\int_0^{\frac{\tau}{2}}\dot k\sigma_xdt=\frac{1}{4}\int_0^{\frac{\tau}{2}}\dot k_1\frac{D_1}{\gamma_1 k_1}dt=-\frac{D_1}{4\gamma_1}\ln 2.
\label{W1}
\eeqa  
Similarly, quasistatic work during compression is given by
\beqa
\langle W_2\rangle =\frac{1}{2}\int_{\frac{\tau}{2}}^{\tau}\dot k\sigma_xdt =\frac{1}{4}\int_{\frac{\tau}{2}}^{\tau}\dot k_2\frac{D_2\tau_a}{k_2(\gamma_2+k_2\tau_a)}dt 
=\frac{D_2\tau_a}{4\gamma_2}\ \ln 2-\frac{D_2\tau_a}{4\gamma_2}\ \ln\left(1+\frac{k_0\tau_a}{2\gamma_2+k_0\tau_a}\right).\nonumber \\
\label{W2}
\eeqa 
So, the total average work along a cycle in quasistatic limit is given by,
\beqa
\langle W\rangle =\langle W_1\rangle +\langle W_2\rangle = -\frac{D_1}{4\gamma_1}\ln 2 +
\frac{D_2\tau_a}{4\gamma_2}\ \ln 2-\frac{D_2\tau_a}{4\gamma_2}\ \ln\left(1+\frac{k_0\tau_a}{2\gamma_2+k_0\tau_a}\right).
\label{W}
\eeqa  
From above expression, first we note that average quasistatic work may not be negative always. If $\frac{D_2\tau_a}{4\gamma_2}\ln 2 < \frac{D_1}{4\gamma_1}\ln 2+\frac{D_2\tau_a}{4\gamma_2}\ln\left(1+\frac{k_0\tau_a}{2\gamma_2+k_0\tau_a}\right)$, then work can be extracted in quasistatic limit from the system. Such conditions can be set up in experiments \cite{Libchab2000, MaggiPRL14}. Though in $\tau_a\rightarrow \infty$  or $\gamma_2\rightarrow \infty$ limit, one can always extract $\frac{D_1}{4\gamma_1}\ln 2$ amount of average thermodynamic work. Therefore in these limiting cases, the system always works as an engine. 
\begin{figure}[!t]
\centering
\includegraphics[width=0.5\columnwidth]{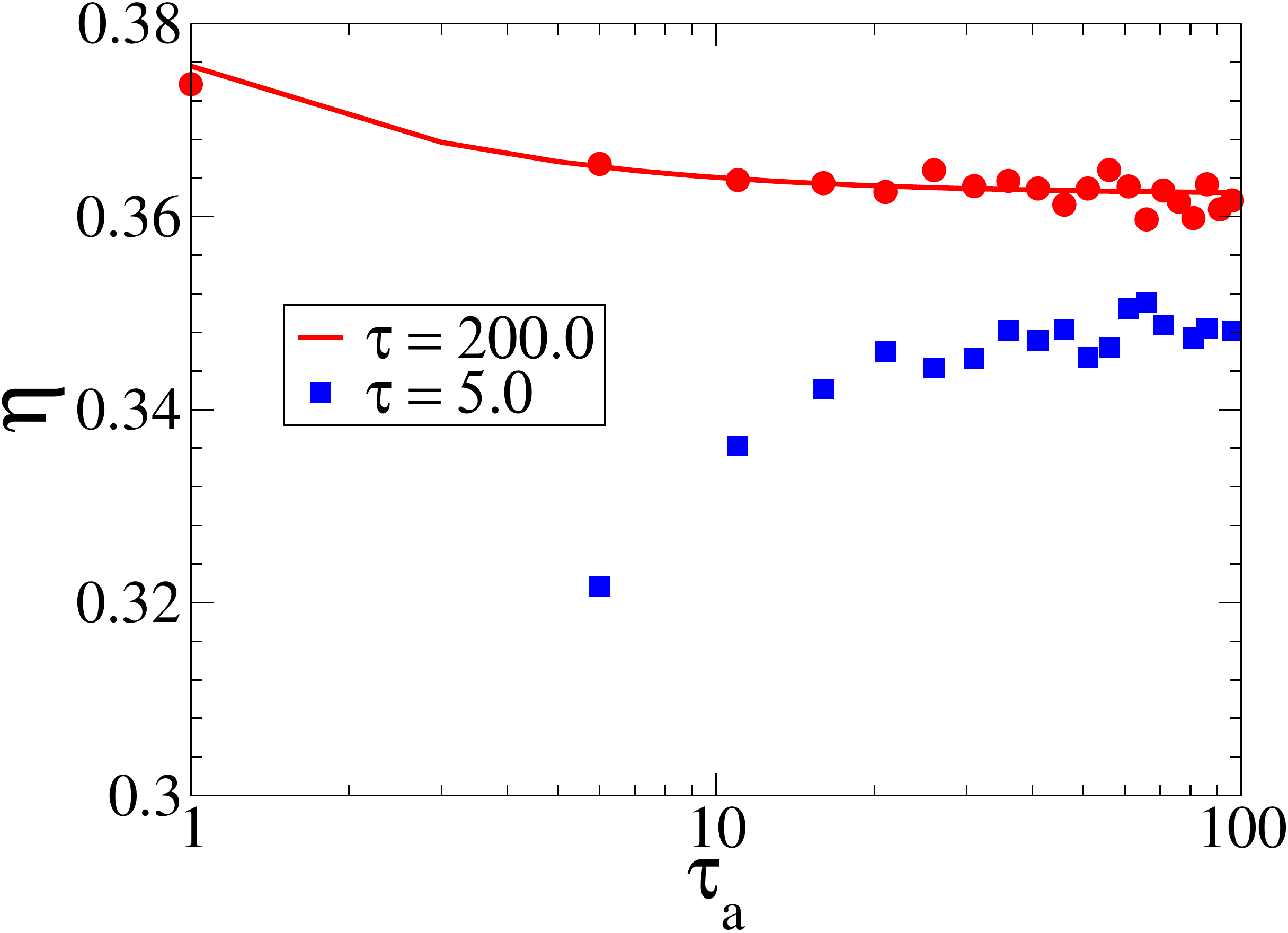}
\caption{Plot of efficiency $\eta$ as function of correlation time $\tau_a$ for the analytical expression in quasistatic case $(\tau=200)$ from Eq. (\ref{geneta}) and simulations (red circles), when compared with the efficiency in nonquasistatic regime $(\tau=5.0)$ from simulations (blue squares). Other parameters are $T=4$, $k_0=5$, $\gamma_1=4$, $\gamma_2=1$, $D_2=5$, $n=2$.}
\label{effigen}
\end{figure}
Now we calculate the average efficiency of the system while working in the engine mode in quasistatic limit. To do that we need to calculate the average heat exchanged ($\langle Q_{in}\rangle $) between the system and the bath during expansion of the trap. From first law we know $\langle Q_{in}\rangle =\Delta U_e - \langle W_1\rangle $ where $\Delta U_e$ is the average change of energy of the particle during the expansion, i.e. $\Delta U_e=\langle U(\frac{\tau}{2}^-)\rangle -\langle U(0^-)\rangle =\frac{D_1}{4\gamma_1}-\frac{D_2\tau_a}{4(\gamma_2 + k_0\tau_a)}$. Therefore, in quasistatic limit $\langle Q_{in}\rangle =\Delta U_e+\frac{D_1}{4\gamma_1}\ln2$ and the average efficiency is given by,        
\beqa
\eta=\frac{\frac{D_1}{4\gamma_1}\ln 2 -
  \frac{D_2\tau_a}{4\gamma_2}\ \ln 2+\frac{D_2\tau_a}{4\gamma_2}\ \ln\left(1+\frac{k_0\tau_a}{2\gamma_2+k_0\tau_a}\right)}{\frac{D_1}{4\gamma_1}\ -\frac{D_2\tau_a}{4(\gamma_2+k_0\tau_a)}\ + \frac{D_1}{4\gamma_1}\ \ln2}\ .
\label{geneta}
\eeqa

\noindent Here we stress that this expression is valid only when $\langle W\rangle < 0$ and $\langle Q_{in} \rangle  > 0$ such that $\eta >0 $. We compare the simulation results
with the analytical expressions for $\eta$ for quasistatic case $\tau=200$ in Fig. (\ref{effigen}) and also the non-quasitatic case from the simulations for $\tau=5$.
One sees that quasistatic case is more efficient than the non-quasistatic case. 

\section{Particular case} \label{part_results}
In this section we look at the particular case where due to activity only the characteristics of fluctuations in the equation of motion alters breaking FDR. However the friction is unaltered. When the bath becomes active, the fluctuations become exponentially time-correlated and their strength is related to the correlation time in a particular way as in \cite{Fodor16, MandalPRL17}, the rationale of which is explained below.    

While working in the engine mode, one can consider the passive limit of the system where the activity of the bath used during $t=\tau/2$ to $t=\tau$, is reduced to zero. In this limit, as the particle is always in contact with a single thermal bath, for any time-periodic as well as quasistatic drive we expect $\langle W\rangle \rightarrow 0$.
From the above expression of $\langle W\rangle$, this is possible only when one considers:  $\gamma_2\rightarrow \gamma_1$, $ D_2 \rightarrow \frac{D_1}{\tau_a}$ and then $\tau_a\rightarrow 0$. This describes the passive limit of heat engine considered here. Furthermore, now if we go back to Eq. (\ref{Lang2}), we note that the fluctuating force in this limit becomes delta correlated and obeys FDR. Because $\lim_{\tau_a\rightarrow 0}{D_2}\langle\xi_2(t)\xi_2(t^{\prime})\rangle=D_1\lim_{\tau_a\rightarrow 0}\frac{1}{2\tau_a}\exp^{-\frac{|t-t^{\prime}|}{\tau_a}}=D_1\delta(t-t^{\prime})$ which is consistent with the physics of thermal bath equilibrated at temperature $T$.
\begin{figure}[!htbp]
\centering
\includegraphics[width=0.32\columnwidth]{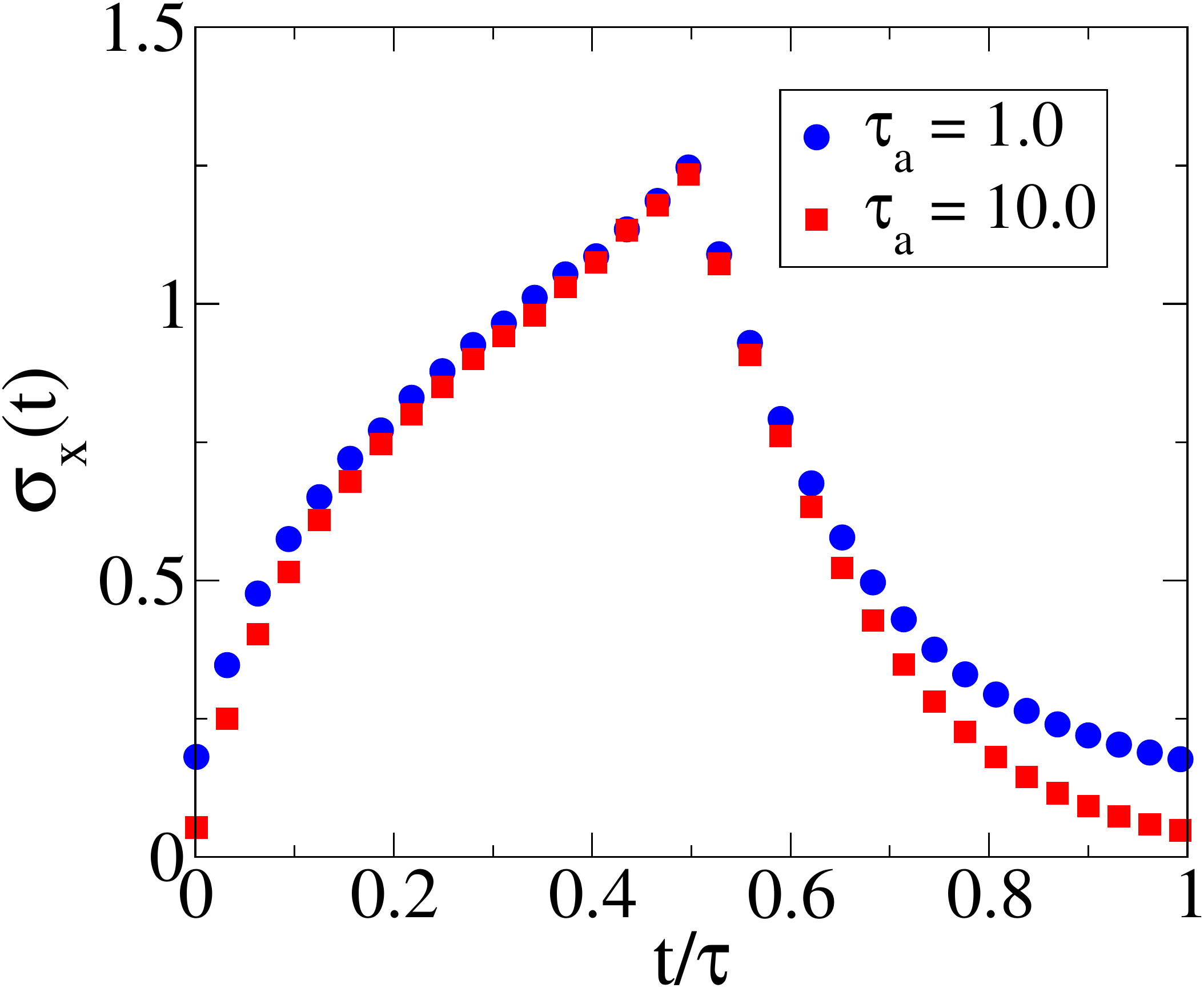}
\includegraphics[width=0.32\columnwidth]{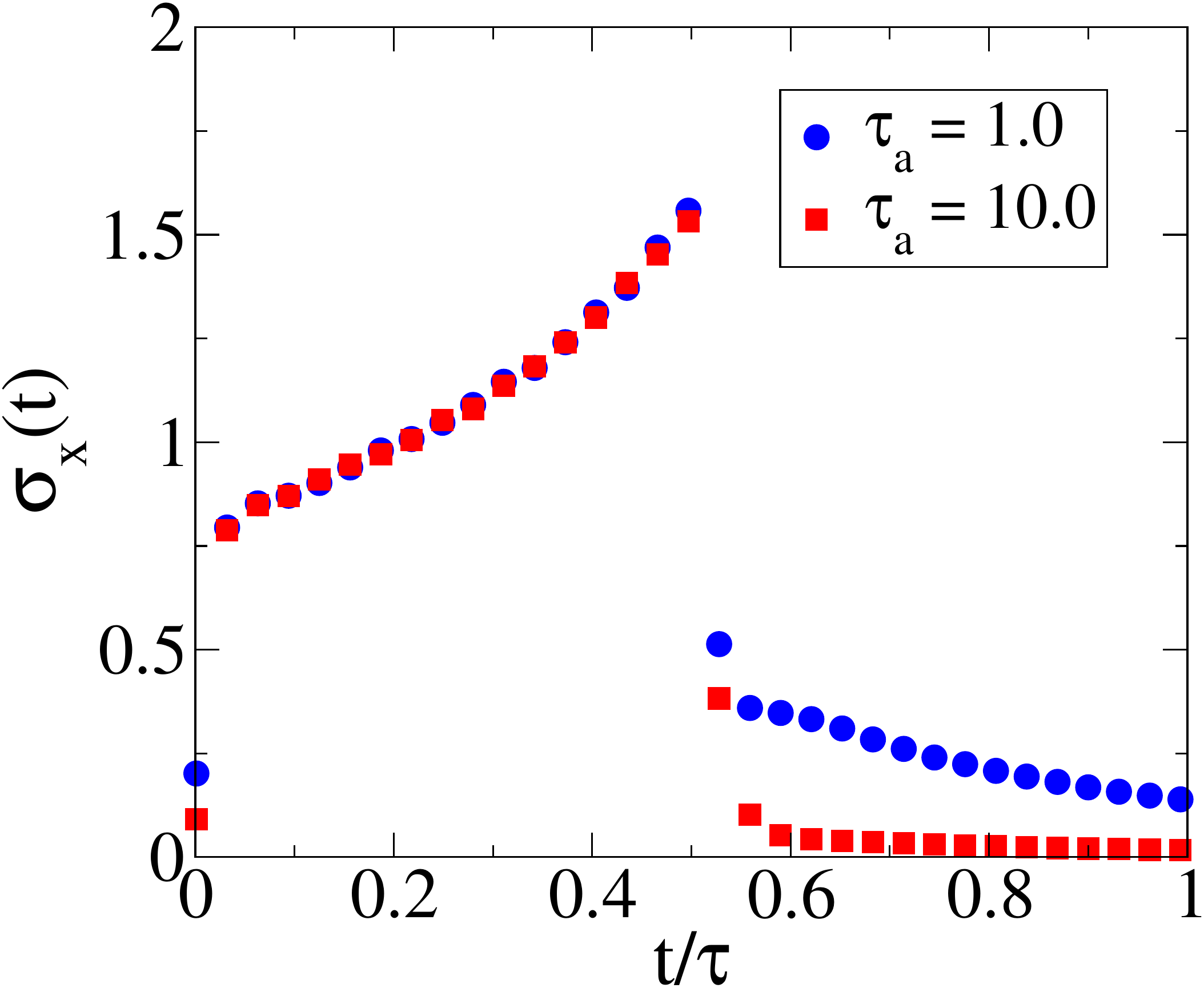}
\includegraphics[width=0.32\columnwidth]{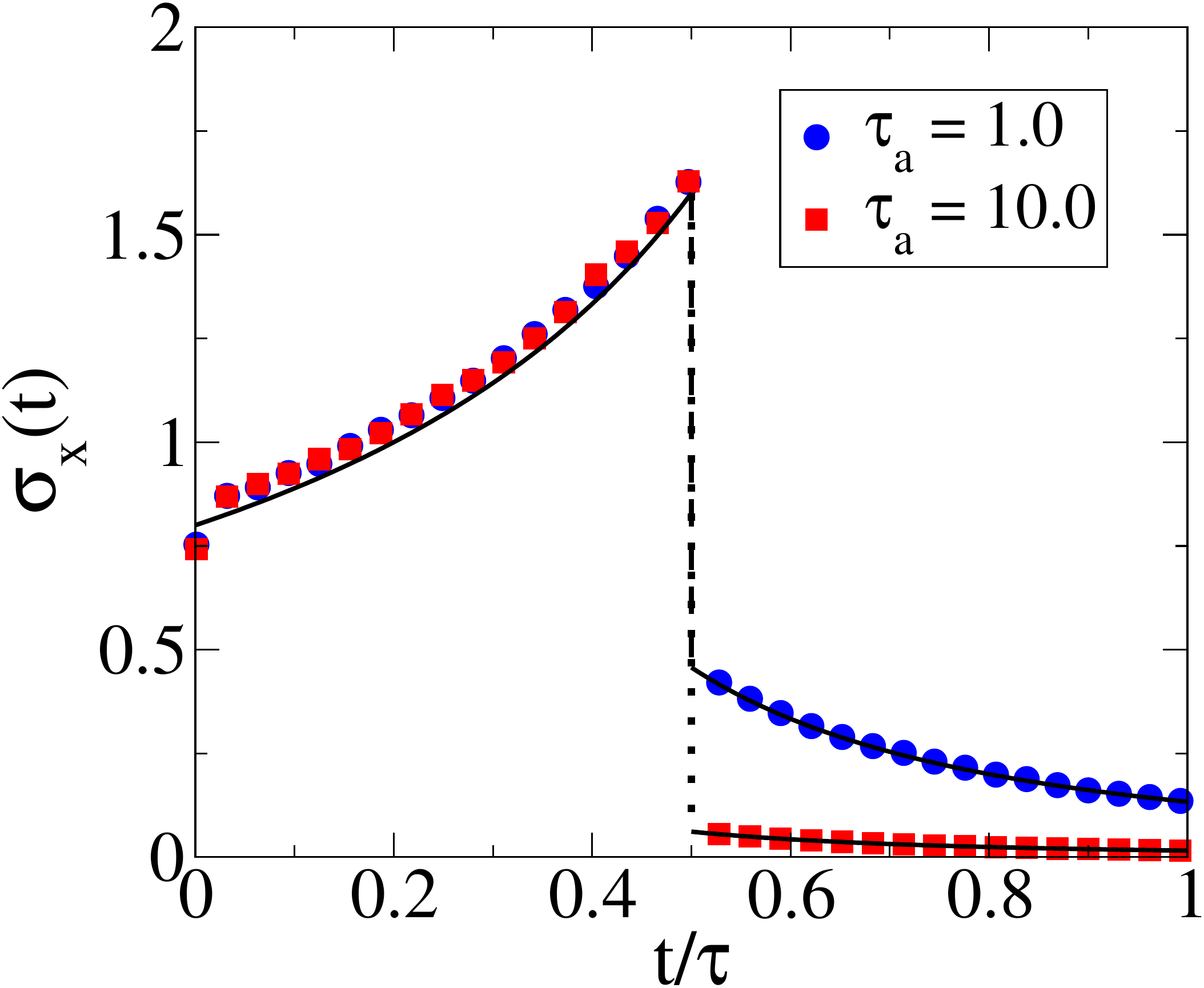}
\caption{Plot of $\sigma_x(t)$ vs $t/\tau$ for different cycle times $\tau=1$ (left panel), $\tau=10$ (middle panel) and $\tau=200$ (right panel) from simulations. Note that for $\tau=200$ (right panel) solid lines correspond to quasistatic analytic expressions in Eqs. (\ref{partsigma}), dashed lines at $t/\tau=0.5$ are the jumps calculated from Eqs. (\ref{sigma}). Other parameters are $T=4$, $k_0=5$, $\gamma_1=\gamma_2=1$, $n=2$ and $ D_2=D_1/\tau_a$.}
\label{sigxt}
\end{figure}
Now, if one considers $D_2\rightarrow \frac{D_1}{\tau_a}$ and then $\tau_a\rightarrow 0$ but $\gamma_2\neq\gamma_1$, the bath is still active. It does not obey FDR and thermodynamic work can be extracted from such a system. Similar system is detailed in \cite{Saha18}. Below we focus on the other combination where $\gamma_2\rightarrow \gamma_1 \equiv \gamma$, $D_2\rightarrow \frac{D_1}{\tau_a}$ but with $\tau_a > 0$ as in \cite{Fodor16, MandalPRL17}, such that the bath during compression is still active and thereby remains out of equilibrium to provide persistent, exponentially correlated noise to the colloidal particle immersed into it.

Below we give analytical expressions for all the thermodynamic quantities of interest in the particular case and match the results with simulations. In simulations we fix $k_0=5$, temperature $T=4$ in terms of the units mentioned earlier. We have also chosen $\gamma_1=\gamma_2 \equiv \gamma$ and $D_2=D_1/\tau_a$ such that the system equilibrates at temperature $T$ when $\tau_a\rightarrow 0$.

\begin{figure}[t]
\centering
\includegraphics[width=0.32\columnwidth]{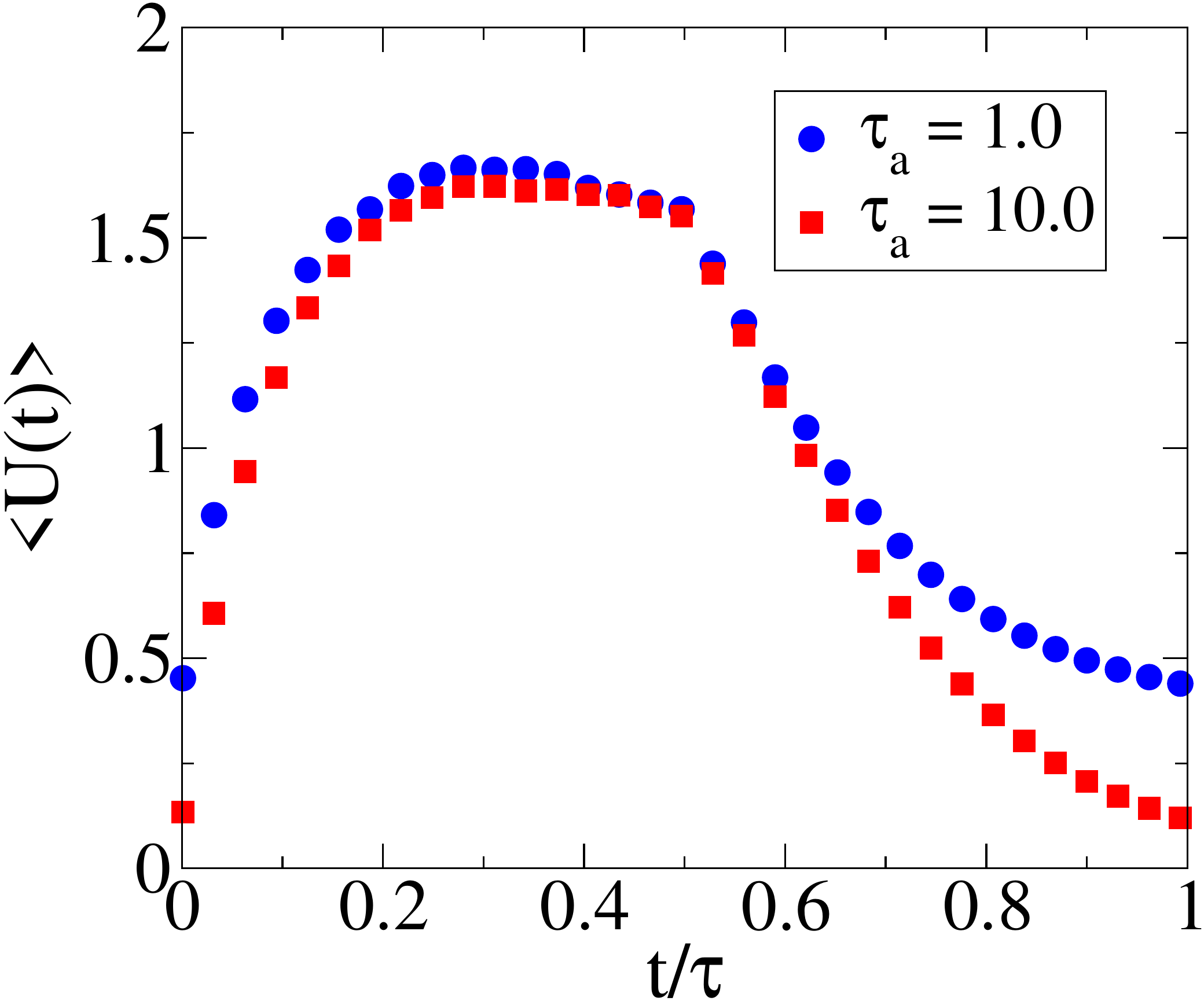}
\includegraphics[width=0.32\columnwidth]{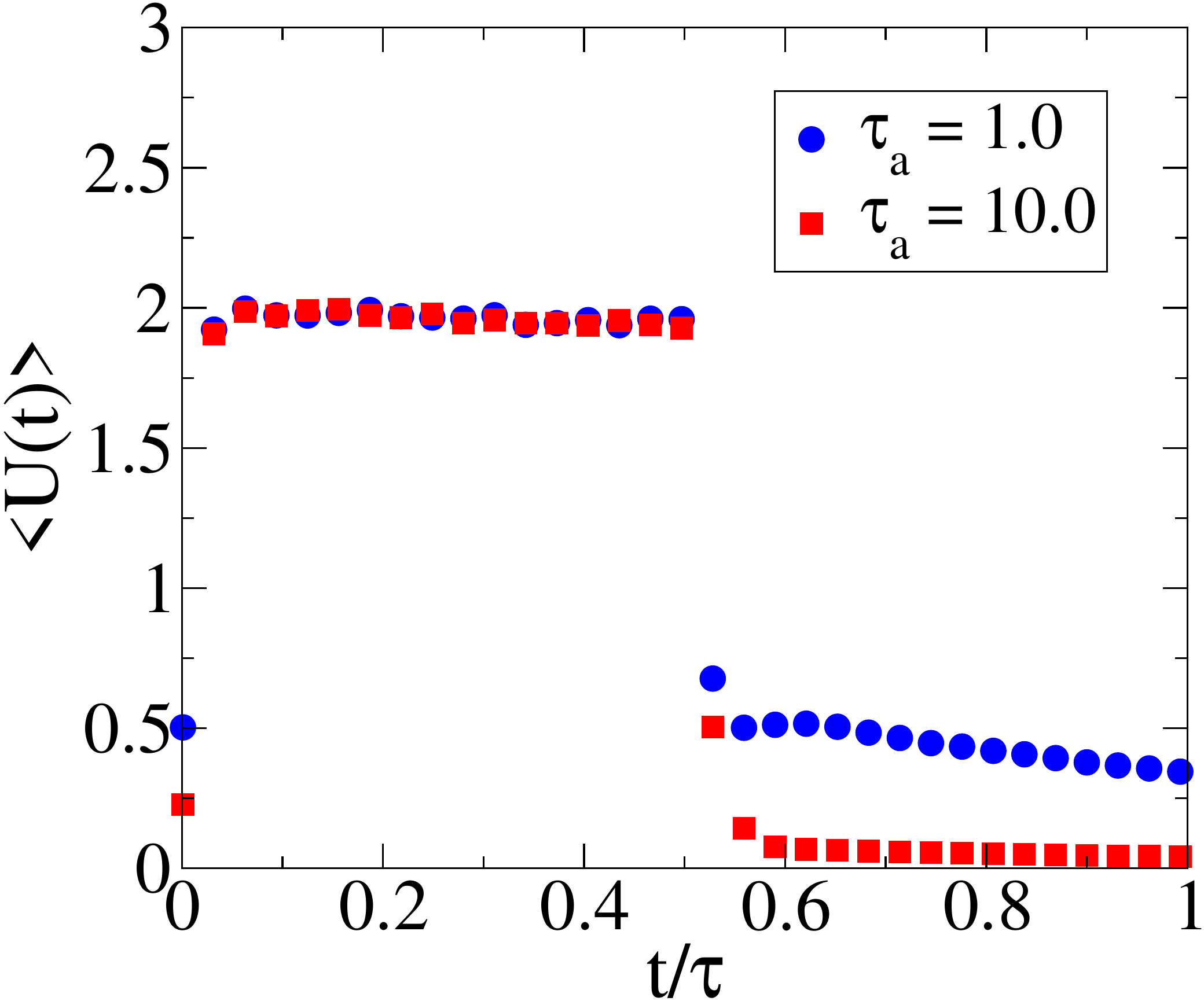}
\includegraphics[width=0.32\columnwidth]{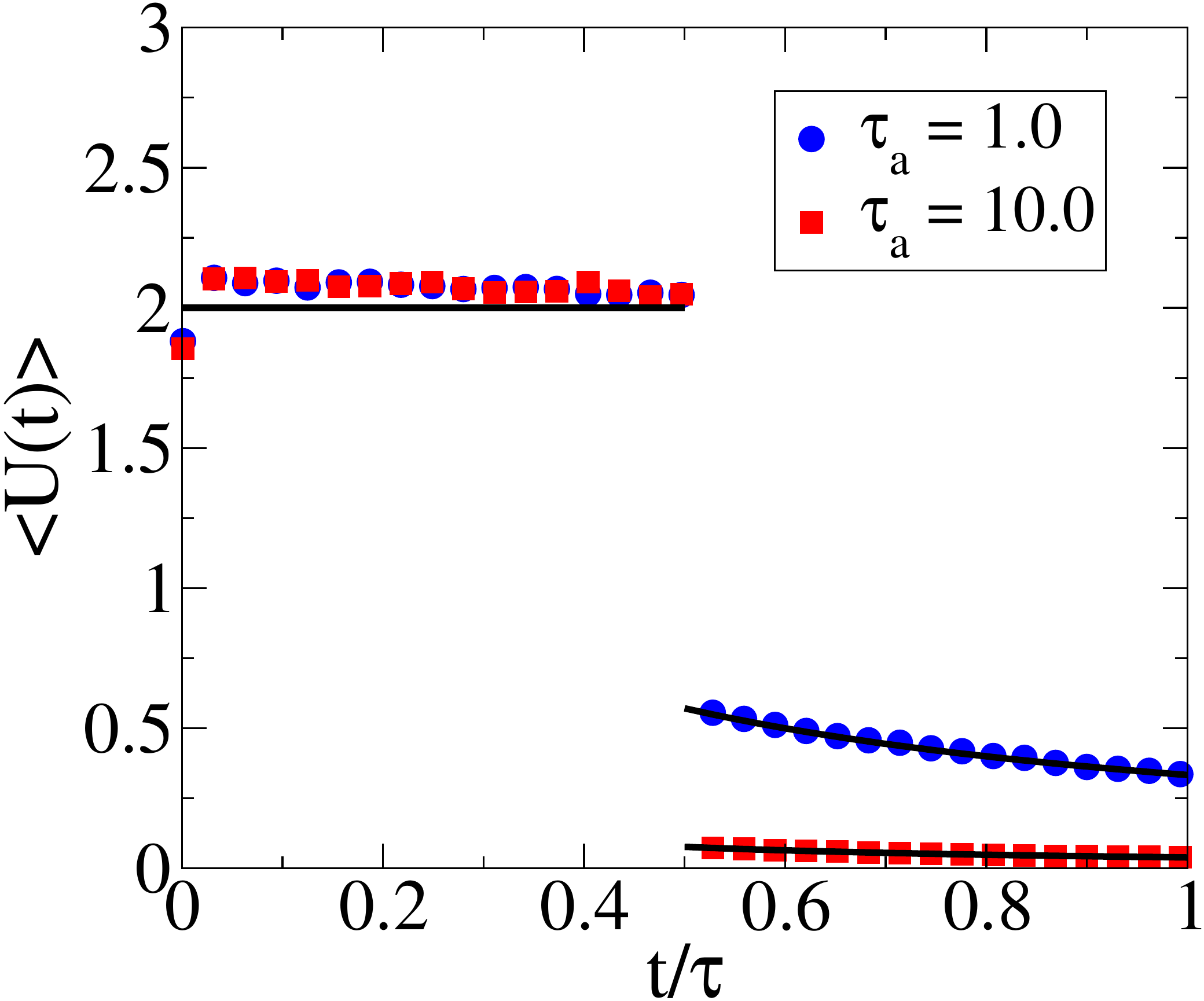}
\caption{Plot of average internal energy $\langle U(t)\rangle $ vs $t/\tau$ for different cycle times $\tau=1$ (left panel), $\tau=10$ middle panel and $\tau=200$ (right panel) from simulations. For $\tau=200$ (right panel) solid lines correspond to quasistatic calculations discussed in the text. Other parameters are $T=4$, $k_0=5$, $\gamma_1=\gamma_2=1$, $n=2$ and $D_2=D_1/\tau_a$.}
\label{energyU}
\end{figure}

In this case the expression for $\sigma_x$ in the quasistatic limit turns out to be,
\beqa
\sigma_x(t) \approx
\begin{cases}
\frac{D_1}{2\gamma k_1(t)}\ , ~~~~~~~~~~~~~~0 < t\leq \tau/2, \\ 
\frac{D_1}{2k_2(t)(\gamma+k_2(t)\tau_a)}\ , ~~~~\tau/2 < t \leq \tau. 
\end{cases}
\label{partsigma}
\eeqa
In figure ({\ref{sigxt}}) we have plotted $\sigma_x$ with $t/\tau$ for two different $\tau$ and $\tau_a$, obtained from simulation. For shorter $\tau$, it does not match with analytical results above but for longer $\tau$ it does. This validates above mentioned analytical expression for $\sigma_x$ in quasistatic limit.
We also plot the average energy $\langle U\rangle$ from simulation for two different $\tau$ and $\tau_a$ and plot it with $t/\tau$ (see figure (\ref{energyU})). We see very good agreement between the analytical and simulation results in the quasistatic limit namely with $\langle U\rangle =\frac{D_1}{4\gamma}$ for expansion step and $\langle U\rangle =\frac{D_1}{4(\gamma+k_2(t)\tau_a)}$ for compression step.
For the special case of parameters, with a little algebra, the average work in Eq. (\ref{W}) can be expressed as,
\[\langle W\rangle =-\frac{k_BT}{2}\ \ln\left(1+\frac{1}{1+X}\right),\] 
where $X=\frac{2\gamma}{k_0\tau_a}$ is defined to be the ratio of two time scales involved in the model namely $1/\gamma$ and $1/k_0\tau_a$. Therefore, when $X\rightarrow 0$, the maximum work that can be extracted is, 
\[\langle W_{max}\rangle =-\frac{k_BT}{2}\ \ln 2, \]
\begin{figure}[b]
\centering
\includegraphics[width=0.45\columnwidth]{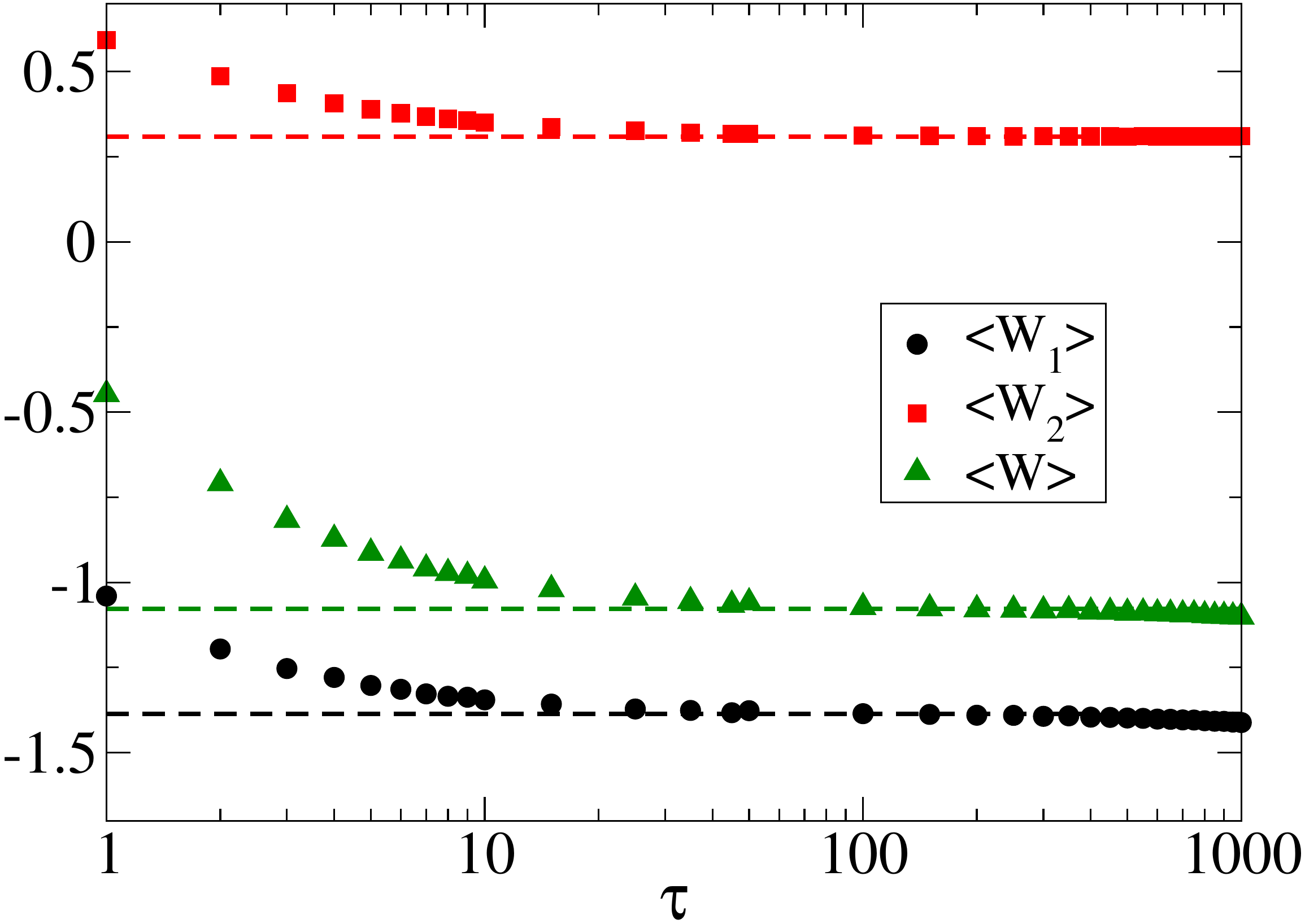}
\includegraphics[width=0.45\columnwidth]{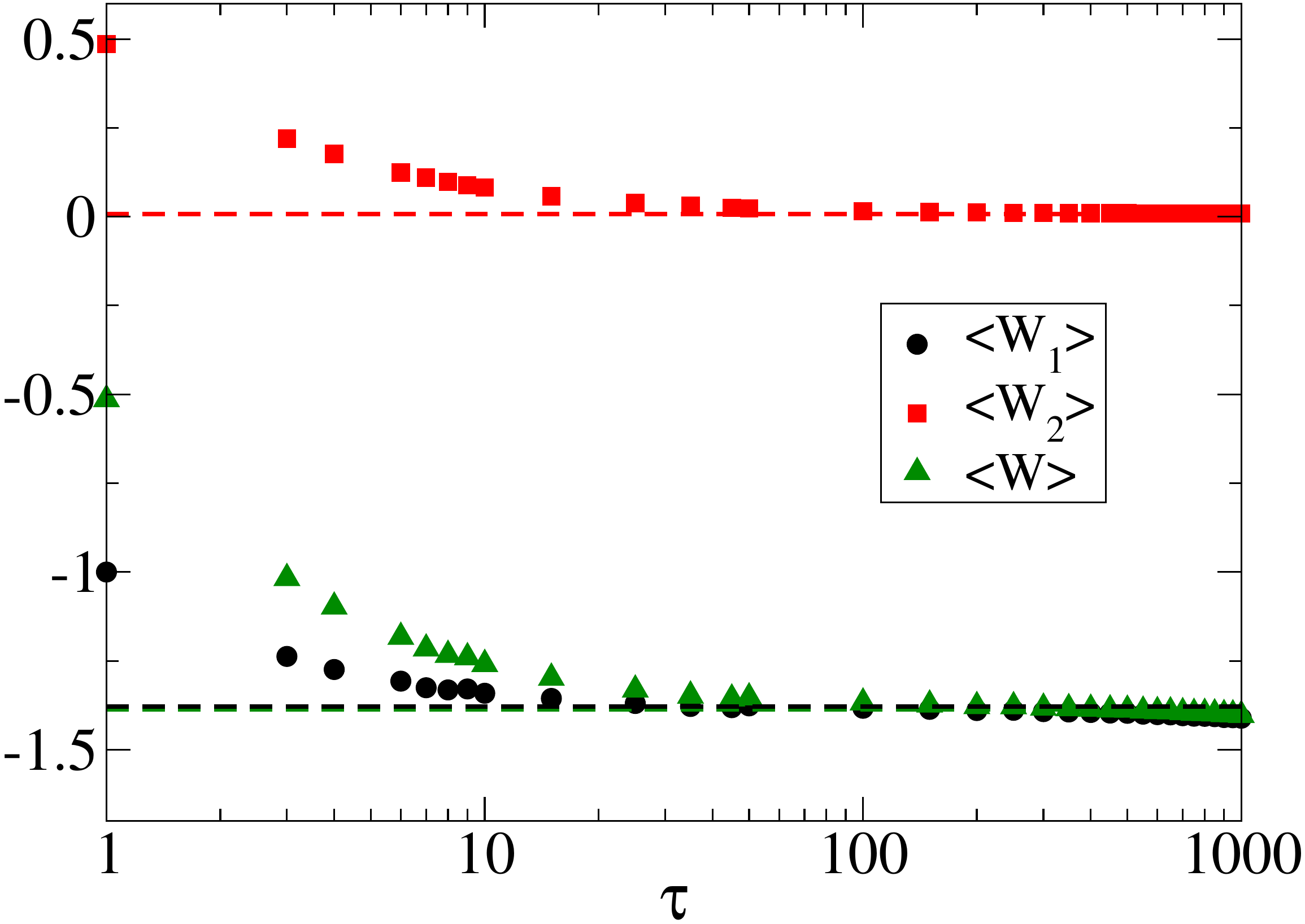}
\caption{Plot of work done $\langle W_1\rangle $, $\langle W_2\rangle $, $\langle W\rangle $ for $\tau_a =1$ (left panel) and $\tau_a =50$ (right panel), as a function of the cycle time $\tau$ from simulations. Dashed lines in both the plots correspond to quasistatic values calculated from Eqs. (\ref{W1}), (\ref{W2}), (\ref{W}) respectively. Other parameters are $T=4$, $k_0=5$, $\gamma_1=\gamma_2=1$, $n=2$ and $ D_2=D_1/\tau_a$.}
\label{Wtau}
\end{figure}
\begin{figure}[t]
\centering
\includegraphics[width=0.45\columnwidth]{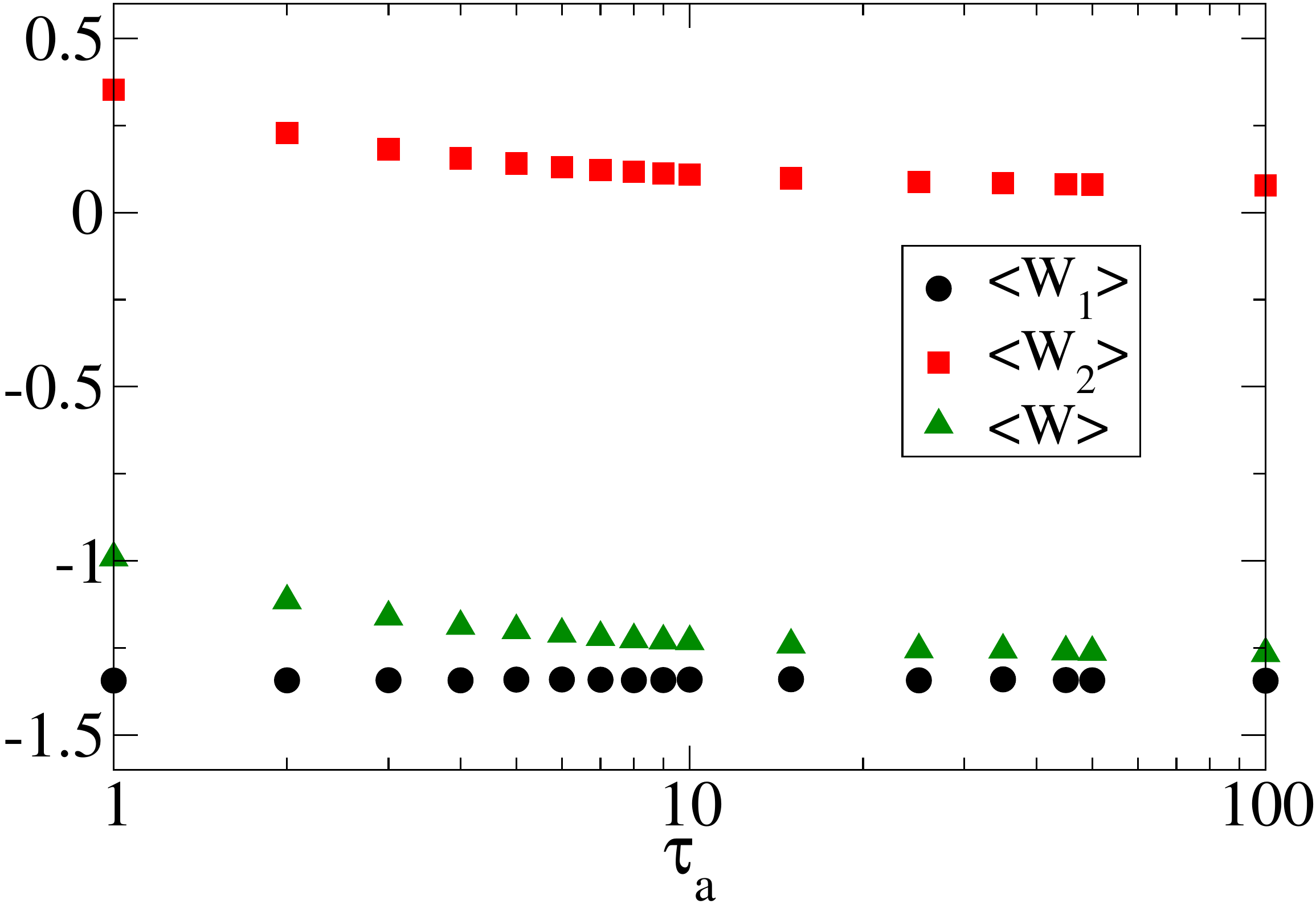}
\includegraphics[width=0.45\columnwidth]{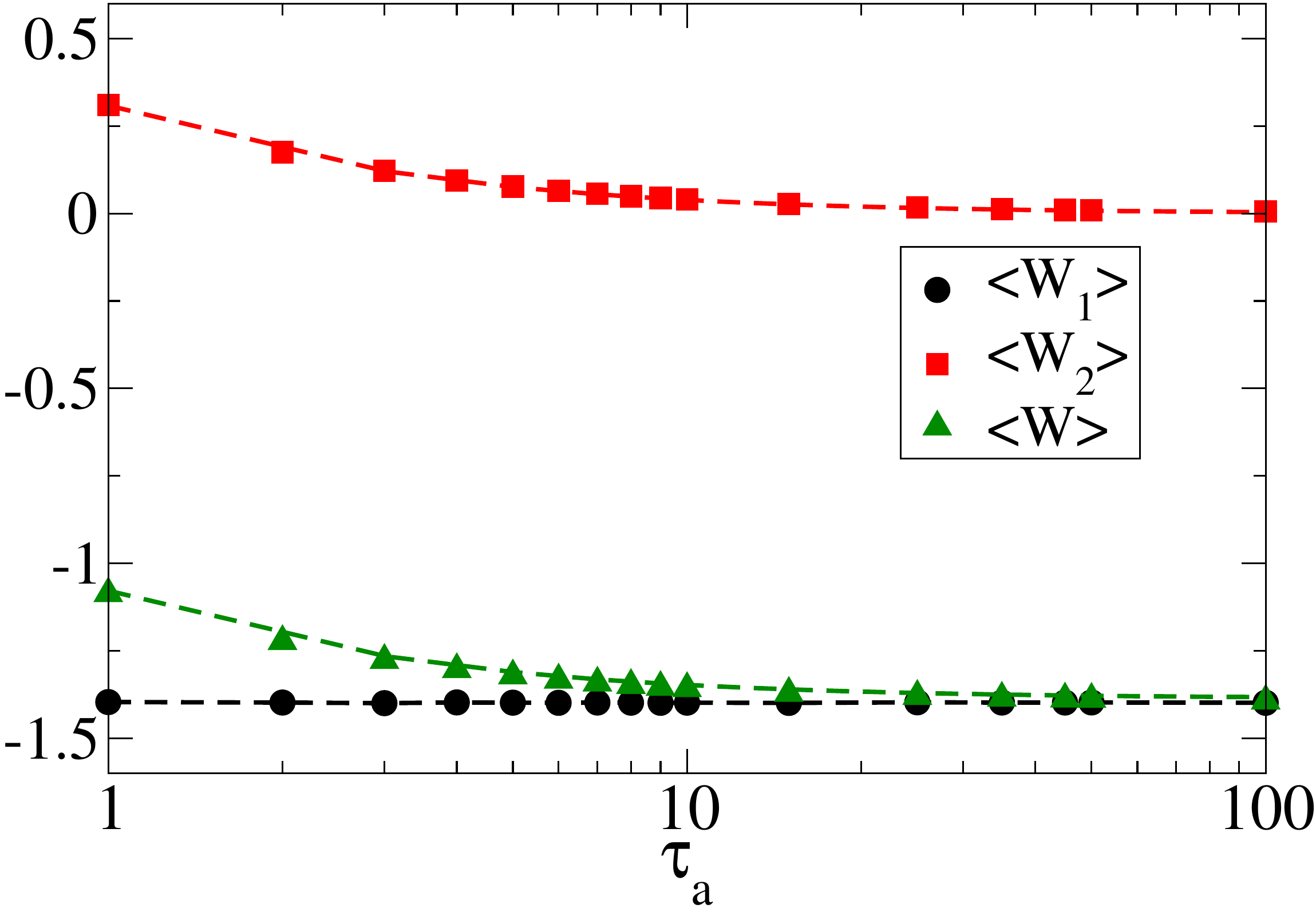}
\caption{Plot of work done $\langle W_1\rangle $, $\langle W_2\rangle $, $\langle W\rangle $
for $\tau =10$ (left panel) and $\tau =500$ (right panel), as a function of the correlation time $\tau_a$ from simulations. Dashed lines in the right panel correspond to quasistatic values calculated from Eqs. (\ref{W1}), (\ref{W2}), (\ref{W}) respectively. Other parameters are $T=4$, $k_0=5$, $\gamma_1=\gamma_2=1$, $n=2$ and $ D_2=D_1/\tau_a$.}
\label{WtauA}
\end{figure}
also note that in the limit $\tau_a\rightarrow 0$ or $X\rightarrow \infty$, total work $\langle W\rangle\rightarrow 0$, thus no work is extracted as expected. This implies that it is possible to extract work from the system only because of the finite (exponential) time correlation between the random forces along the compression step, where AOUP is assumed. In figure (\ref{Wtau}) we have plotted work with cycle time $\tau$ for two different $\tau_a$, from simulation and we have shown that as $\tau$ becomes larger, $W$ approaches towards its quasistatic values that is $\langle W\rangle \rightarrow \langle W_{max}\rangle = -2.0~ \ln 2\sim -1.39$, with $T=4$ and $k_B=1$, as discussed above.

\begin{figure}[t]
\centering
\includegraphics[width=0.32\columnwidth]{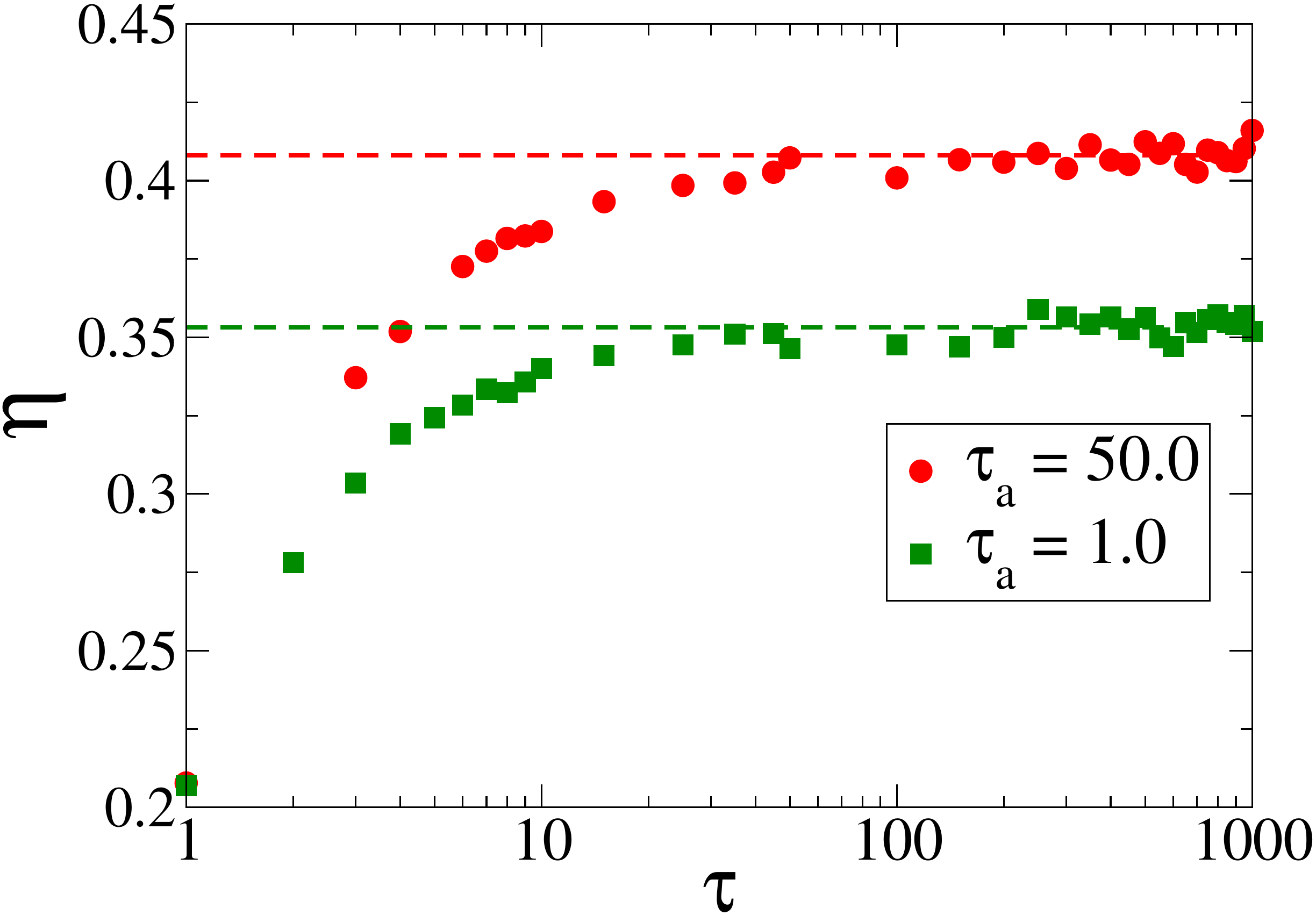}
\includegraphics[width=0.32\columnwidth]{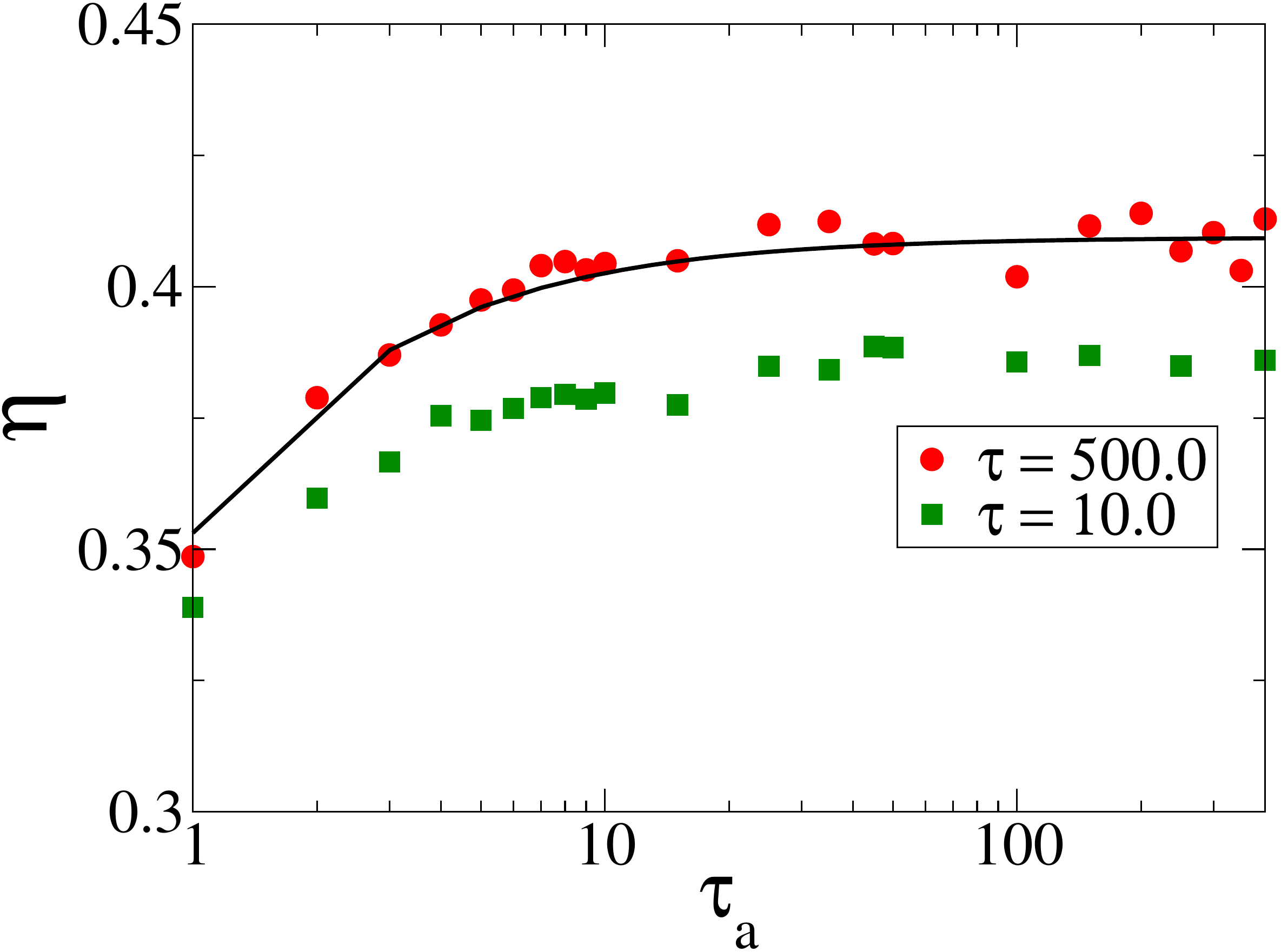}
\includegraphics[width=0.32\columnwidth]{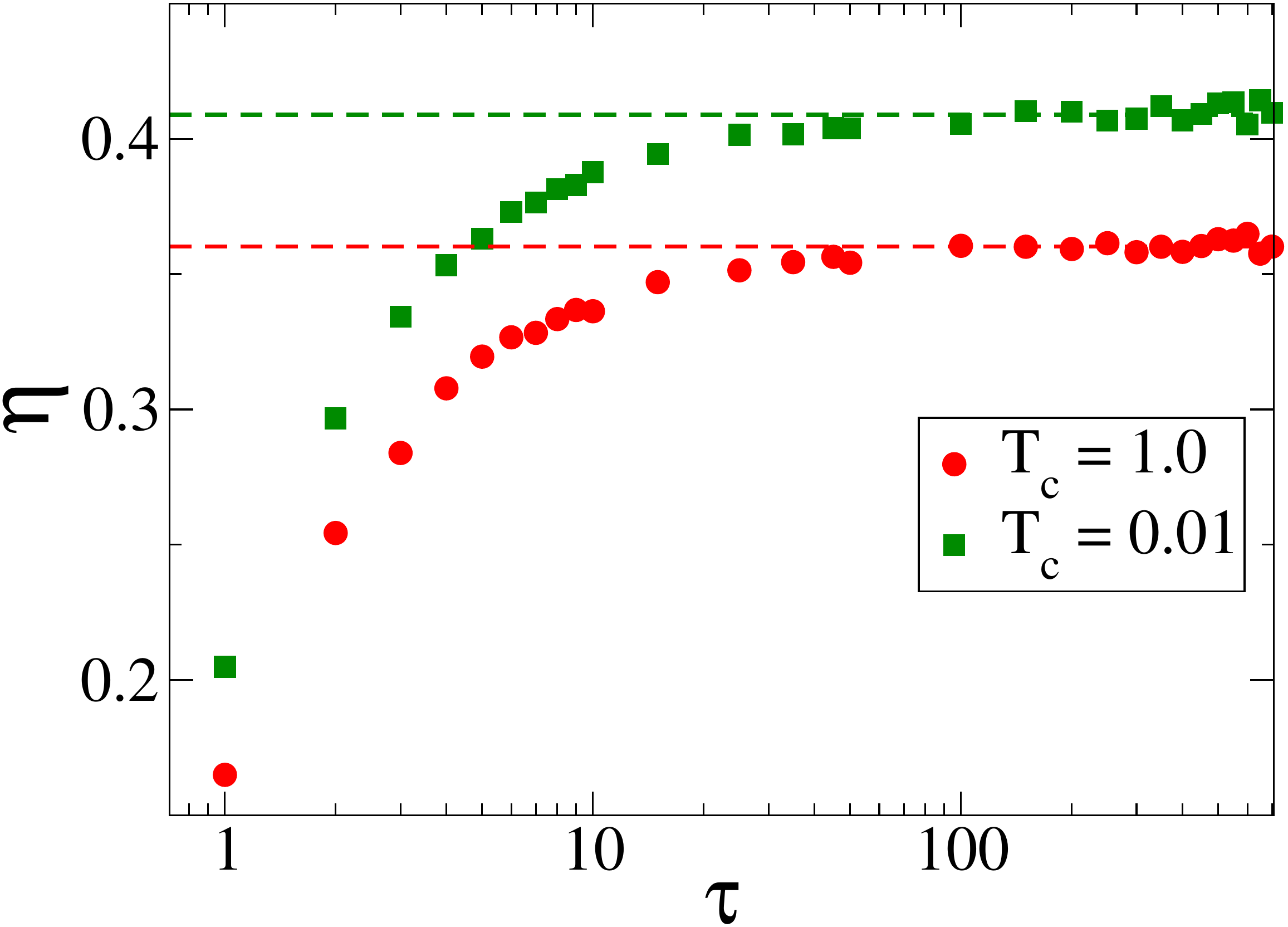}
\caption{Plot of efficiency $\eta$ as function of cycle time $\tau$ (left panel), correlation time $\tau_a$ (middle panel) when compared with the efficiency of the passive engine (right panel) from simulations. Dashed lines in left and right panels correspond to quasistatic analytical expressions Eqs. (\ref{effi1}) and (\ref{effi2}) respectively. $T_c$ in the right panel is the cold bath temperature in case of the passive micro-heat engine. Solid line in the middle panel correspond to Eq. (\ref{effiAA}). Other parameters are $T=4$, $k_0=5$, $\gamma_1=\gamma_2=1$, $n=2$ and $ D_2=D_1/\tau_a$.}
\label{effiA}
\end{figure}

In figure (\ref{WtauA}) we have plotted $\langle W\rangle$ with correlation time scale $\tau_a$ for two different $\tau$. In case of longer cycle times, the work extracted for various $\tau_a$ matches with above mentioned analytical results in quasistatic limit. We now move on to discuss the efficiency. In the special case it becomes,
\beqa
\eta=\frac{\ln\left(1+\frac{k_0\tau_a}{2\gamma+k_0\tau_a}\right)}{\ln2+\frac{k_0\tau_a}{\gamma+k_0\tau_a}}.
\label{effiAA}
\eeqa
The efficiency can be rewritten as $\eta=\frac{\ln(1+\frac{1}{1+X}\ )}{\ln2+\frac{2}{2+X}}$. In the limit $X\rightarrow 0$ (which can be achieved by taking large $\tau_a$),  $\eta \rightarrow \frac{\ln 2}{1+\ln 2}\simeq 0.41 = \eta_{max}$. Again, when $X\rightarrow \infty$ (which can be achieved by taking very small $\tau_a$), and therefore, $\eta \rightarrow \eta_{min}=0$. 
Therefore the system can produce thermodynamic work and be efficient up to a certain extent only due to finite, nonzero noise correlation specified by AOUP, along compression of the trap. From simulation we compute average efficiency for various $\tau$ and $\tau_a$. In figure (\ref{effiA}) we have plotted average efficiency with $\tau$ for two different $\tau_a$, where for longer $\tau$ it matches with quasistatic efficiency, calculated before. In the same figure (different panel), we have plotted average efficiency with $\tau_a$ for two different $\tau$, where the efficiency for longer $\tau$ matches quite well with its quasistatic value.   
Next we will compare the micro heat engine of our concern with another similar micro heat engine in terms of efficiency, where the only difference between these two systems is: during compression the trapped particle is immersed in a thermal bath of temperature $T_c < T$ instead of an active bath. This system has been realized experimentally, as Stirling-type micro heat engine \cite{bechinger10}. The time-periodic protocol here is varied linearly between $k_{max}=k_0$ and $k_{min}=\frac{k_0}{n}$ ($n \ge 2$) with the cycle time $\tau$ as prescribed in Eq. (\ref{protocol1}). We will compare the efficiencies of these two micro heat engines with increasing $n$. The average efficiency given in Eq. (\ref{effiAA}) can be generalized with $n$ as 
\beqa
\eta_1(Z_1,n)= \frac{\ln\left(1+\frac{(n-1)Z_1}{n+Z_1}\right)}{\ln n+\frac{Z_1}{1+Z_1}},
\label{effi1}             
\eeqa   
where $Z_1=\frac{\tau_ak_0}{\gamma}$. Similarly, the average quasistatic efficiency of the Stirling-type micro heat engine working between two thermal baths of temperatures $T$ and $T_c$ ($T>T_c$) is given by \cite{bechinger10, Sood16},
\begin{figure}[t]
\centering
\includegraphics[width=0.5\columnwidth]{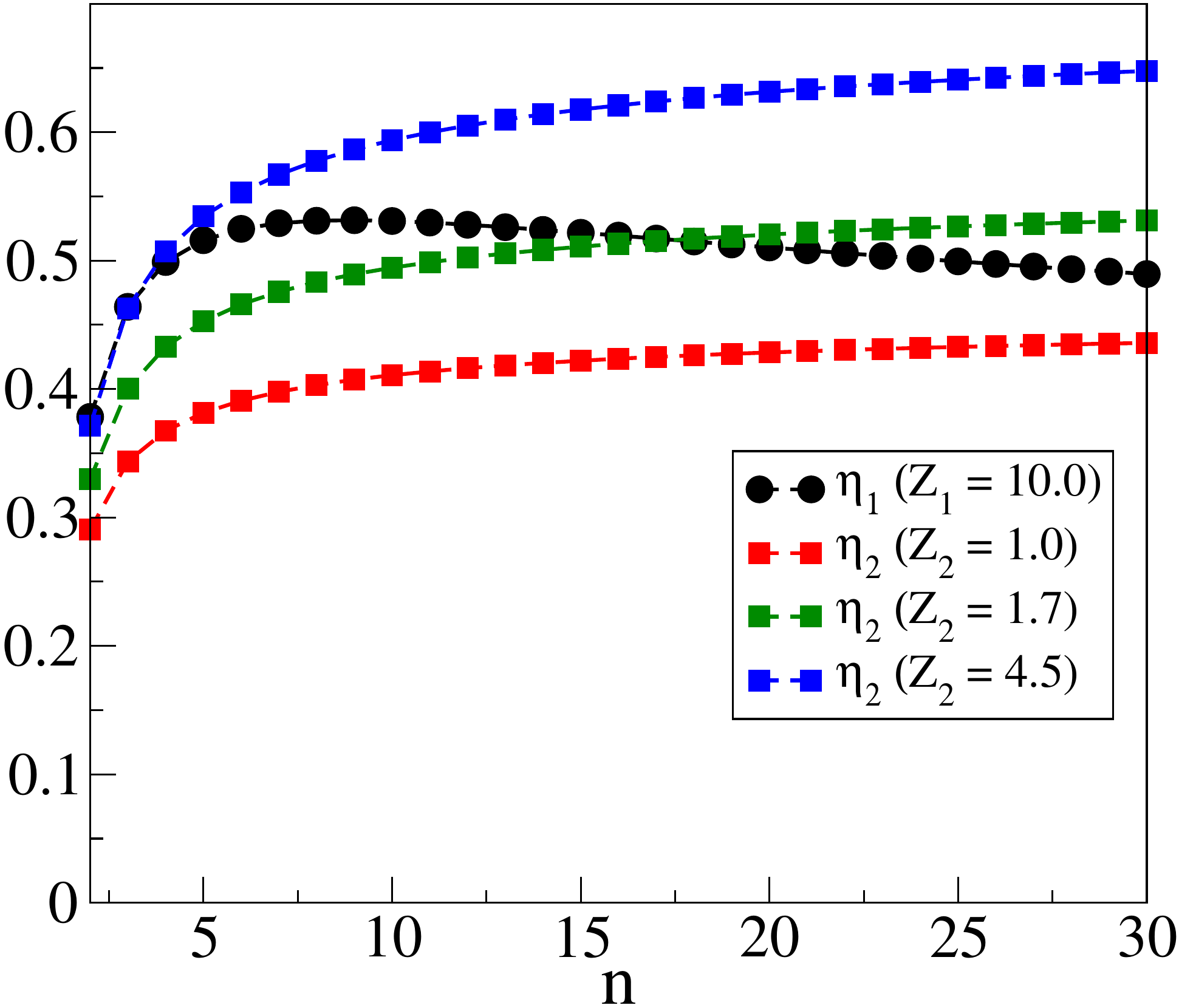}
\caption{Comparison of active and passive efficiencies as a function of $n$ from Eq. (\ref{effi1}) and (\ref{effi2}). One can see that the active engine surpasses the passive engine's efficiency for certain parameter regime.}
\label{compare}
\vspace{-3mm}
\end{figure}  

\beqa
\eta_2(Z_2,n)&=&\frac{Z_2\ln n}{Z_2+(Z_2+1)\ln n},
\label{effi2}
\eeqa 
where $Z_2=\frac{T}{T_c}-1$. Note that here, $0<Z_j<\infty$ and $1<n<\infty$ where $j\in (1,2)$. Again, $\lim_{Z_j\rightarrow 0}\eta_j=0$ and $\lim_{Z_j\rightarrow\infty}\eta_j=\frac{\ln n}{1+\ln n}$, which implies they both reach same limiting value in large $Z_j$ limit, keeping $n$ constant. In figure (\ref{compare}) we plot $\eta_j$ with respect to $n$. For $j=1$, $Z_j=10$ and for $j=2$, we consider three different values of $Z_j$ ($=1.0, 1.7, 4.5$) so that one can easily compare $\eta_1$ and $\eta_2$. From the figure it is clear that $\eta_1$ is a non-monotonic function of $n$ for a given $Z_1$ whereas $\eta_2$ is a monotonically increasing function of $n$. Depending on the values of $Z_2$ and $n$, $\eta_2$ can be lower or higher than $\eta_1$. Therefore, in general $\eta_1$ cannot be always higher or lower than $\eta_2$. It depends on the chosen parameter space. 

For the model systems we consider here, the efficiencies depend on the ratio of the highest and lowest value of the time dependent protocol (i.e. the time dependent spring constant of the harmonic trap) and the values of the control parameters of the respective engines i.e. in case of passive engine it is the ratio of hot and cold bath temperatures and in case of active-bath heat engines it is primarily the correlation time involved in AOUP required to model the effect of activity.      

\section{Conclusion} \label{conclude}
Here we have considered a model of micro heat engine that involves a harmonically trapped colloidal particle driven between a thermal (equilibrated at temperature $T$) and an athermal, out-of-equilibrium bath, by the time-periodic spring constant of the trap. The dynamics of the particle is modeled by overdamped Langevin equations with Gaussian, delta-correlated noise that maintains FDR, when it is in contact with the thermal bath and with exponentially correlated, FDR-breaking noise when subject to the athermal bath. The trap expands linearly with time when the particle is in thermal bath and it contracts, again linearly, when it is in athermal bath. The expansion and contraction is time-periodic as well as time-symmetric (i.e. the expansion and contraction both run exactly up to half of a full cycle). The protocol to drive the engine (i.e. the colloidal particle here) is Stirling-type and can be experimentally implemented \cite{Sood16}. 
Overdamped Langevin dynamics with exponentially correlated noise that breaks FDR (namely, AOUP), is often used to model persistent motility of active, self-propelling particles such as bacteria, active colloids \cite{Libchab2000} etc. The dynamics, while used to simulate many-particle-systems, can exhibit collective phenomena that resembles the features observed in active systems \cite{Fodor16}.  Here we have used it to model the dynamics of the colloidal bead in the athermal bath involving a population of self-propelling entities such as bacteria. It has been shown earlier that a colloidal bead in contact with a bacterial bath exhibits short-time super diffusion and long-time normal diffusion which has been successfully modeled by AOUP when used as the equation of motion of the bead \cite{Zakine17}. Our aim here was to show that the colloidal particle when time-periodically switches its dynamics between  thermal, overdamped, Langevin equation and AOUP, in the quasistatic limit under proper conditions (as mentioned in the text before) it can produce thermodynamic work. More over, depending on the parameter space, it can be more efficient than its passive counterpart which is a colloidal particle driven by the same Stirling protocol but within two thermal bath having two different temperatures \cite{bechinger10}. In particular, suppose that the value of correlation time and the friction coefficient of the harmonically trapped colloidal particle following AOUP together with the maximum value of the spring constant of the trap are given. Now, if the ratio between the maximum and minimum value of the spring constant and the ratio between the hot and cold temperatures of the passive micro Stirling engine both are below than a certain threshold value, we have found that on an average passive engine is less efficient than its active counterpart. 

\section{Acknowledgments}
AS thanks University Grants Commission Faculty Recharge Program (UGCFRP), India. Authors also thank E. Roldan for careful reading of the manuscript.


\end{document}